\documentclass[notitlepage,11pt,eqsecnum,amsmath,preprintnumbers,superscriptaddress,nofootinbib,aps]%,twocolumn,10pt]
{revtex4-1}

\usepackage{hyperref}
\usepackage{multirow}
\usepackage{graphicx}
\usepackage{amsmath}
\usepackage{amssymb}
\usepackage{subfigure}
\usepackage{color,bm}

\long\def\del #1 \enddel { }

\usepackage{xcolor}
\usepackage{colortbl}

\definecolor{Gray}{gray}{0.85}
\definecolor{LightGray}{gray}{0.93}
\definecolor{LightGreen}{rgb}{0.88, 1, 0.88}
\definecolor{LightCyan}{rgb}{0.88,1,1}
\definecolor{LightRed}{rgb}{1, 0.85, 0.85}
\definecolor{LightYellow}{rgb}{1, 1, 0.85}
\definecolor{LightBlue}{rgb}{0.87, 0.94, 1}
\definecolor{white}{gray}{1}

\newcolumntype{G}{>{\columncolor{LightGray}}c}
\newcolumntype{L}{>{\columncolor{LightGray}}l}

\renewcommand{\thesection}{{\bf \Roman{section}}}

\usepackage{graphicx}
\usepackage{amsmath}
\usepackage{amssymb}
\usepackage{subfigure}

\usepackage{epsfig}
\def\vev{\rho_0}
\def\mass{m^2}
\def\beq{\begin{equation}}
\def\eeq{\end{equation}}
\def\bea{\arraycolsep .1em \begin{eqnarray}}
\def\eea{\end{eqnarray}}
\def\Tr{{\rm Tr}}

\def\eq#1{(\ref{#1})}

\def\s0#1#2{\mbox{\small{$ \frac{#1}{#2} $}}}
\def\0#1#2{\frac{#1}{#2}}

\usepackage{array,mathtools,amssymb,booktabs}
\newcolumntype{C}{>{$}c<{$}}
\AtBeginDocument{
\heavyrulewidth=.16em
\lightrulewidth=.1em
\cmidrulewidth=.03em
\belowrulesep=.4ex
\belowbottomsep=0pt
\aboverulesep=.4ex
\abovetopsep=0pt
\cmidrulesep=\doublerulesep
\cmidrulekern=.5em
\defaultaddspace=.5em
}
\makeatletter
    \def\CT@@do@color{%
      \global\let\CT@do@color\relax
            \@tempdima\wd\z@
            \advance\@tempdima\@tempdimb
            \advance\@tempdima\@tempdimc
    \advance\@tempdimb\tabcolsep
    \advance\@tempdimc\tabcolsep
    \advance\@tempdima2\tabcolsep
            \kern-\@tempdimb
            \leaders\vrule
                    \hskip\@tempdima\@plus  1fill
            \kern-\@tempdimc
            \hskip-\wd\z@ \@plus -1fill }

   \makeatother
   
\begin{document}

\title{${}$\\[-10ex] Critical $O(N)$ models in the complex field plane}

\author{Daniel F. Litim} 
\author{Edouard Marchais} 
\address{
 Department of Physics and Astronomy, University of
 Sussex, 
BN1 9QH, U.K.}

\begin{abstract}
Local and global scaling solutions for $O(N)$ symmetric scalar field theories are studied in the complexified field plane with the help of the renormalisation group. Using expansions of the effective action about small, large, and purely imaginary fields, we obtain and solve exact recursion relations for all couplings and determine the $3d$ Wilson-Fisher fixed point analytically. For all $O(N)$ universality classes, we further establish that Wilson-Fisher fixed point solutions display singularities in the complex field plane,
which dictate the radius of convergence for real-field expansions of the  effective action.
At infinite $N$, we find closed expressions for the convergence-limiting singularities and prove that local expansions of the effective action are powerful enough to uniquely determine the global Wilson-Fisher fixed point for any value of the fields. Implications of our findings for interacting fixed points in more complicated theories are indicated. 
\end{abstract}
 \vskip-.6cm

\maketitle

\tableofcontents

\newpage
\section{\bf Introduction}

Fixed points play an important r$\hat{\rm o}$le in quantum field theory and statistical physics.  Infrared (IR) fixed points  characterise the low-energy behaviour of theories including  continuous phase transitions or the dynamical breaking of symmetry  \cite{Wilson:1971bg,Wilson:1971dh,ZinnJustin:1996cy}.  Ultraviolet (UV) fixed points are crucial for the predictivity of theories up to highest energies such as in asymptotic freedom \cite{Gross:1973id,Politzer:1973fx} or asymptotic safety 
\cite{Weinberg:1980gg,Litim:2014uca,Bond:2016dvk,Litim:2011cp}. Correlation functions become independent of length or momentum scales in the vicinity of fixed points, and theories are governed by scaling laws and universal numbers. 
A powerful continuum method to study fixed points  
 is offered by Wilson's renormalisation group (RG) ~\cite{Wilson:1973jj}.
It provides equations for running couplings and $N$-point functions following the successive integrating-out of momentum modes 
from a path integral representation of the theory. 
Various incarnations of the (exact) renormalisation group are available 
\cite{Polchinski:1983gv,Wetterich:1992yh,Morris:1993qb,Litim:2001up,Litim:2000ci}, which, combined with systematic approximation schemes \cite{Golner:1985fg,Litim:1998nf,Berges:2000ew,Litim:2002ce,Pawlowski:2003hq,Blaizot:2005xy,Litim:2010tt},  give access to the relevant physics without being tied to weak coupling. Recent applications cover   theories in fractal or higher dimensions
\cite{Codello:2012ec,Codello:2014yfa,Percacci:2014tfa,Mati:2014xma,Eichhorn:2016hdi,Kamikado:2016dvw}, 
multi-critical phenomena \cite{Eichhorn:2013zza}, supersymmetric models \cite{Synatschke:2010jn,Litim:2011bf,Heilmann:2012yf}, and quantum gravity \cite{Litim:2003vp,Benedetti:2012dx,Demmel:2012ub,Dietz:2012ic,Falls:2013bv,Falls:2014tra}.

Many systems  with interacting fixed points are too complex to allow for exact solutions.
One then has to resort to reliable and tractable approximations such as gradient or vertex expansions. Polynomial approximations of the effective action, however,  have their own limitations: They may lead to spurious fixed points \cite{Margaritis:1987hv,Morris:1994ki,Aoki:1996fn,Aoki:1998um,Litim:2000ci,Litim:2002cf,Defenu:2014jfa} at finite orders,  scaling exponents may \cite{Litim:2002cf} or may not \cite{Aoki:1996fn,Aoki:1998um}  converge to finite limits, and
they invariably have a finite radius of convergence dictated by  
nearby singularities in the complexified field plane \cite{Morris:1994ki,Heilmann:2012yf}. On the other hand, it has also been noted that suitable choices of the wilsonian momentum cutoff can improve the stability and convergence of polynomial expansions
\cite{Litim:2000ci,Litim:2001up,Litim:2001fd}, particularly within the derivative expansion \cite{Litim:2001dt}. Semi-analytical ideas to overcome these shortcomings
have been put forward in \cite{Bervillier:2007rc,Bervillier:2007tc,Bervillier:2008an,Abbasbandy:2011ij} with the help of resummations or conformal mappings. Numerical techniques aimed at high accuracy solutions of functional flows have been developed in 
\cite{Adams:1995cv,Bervillier:2007rc,Borchardt:2015rxa,Borchardt:2016pif}.

In this paper, we are interested in the relation between  {\it local} information about fixed points, extracted from the effective action in a narrow window of field values,
and {\it global} fixed point solutions of the theory, valid for all fields including asymptotically large ones. 
An ideal testing ground for our purposes is provided by self-interacting $O(N)$ symmetric scalar field theories in three  dimensions. 
For all $O(N)$ universality classes, we systematically study the fixed point effective actions for small, large, real, and purely imaginary fields.  In each of these cases, this will provide us with exact recursive relations for all fixed point couplings. We discuss their solutions and conditions under which local expansions are sufficent to access the global fixed point, for all fields. 
We pay particular attention to the occurrence of singularities of the effective action or its derivatives in the complexified field plane, and how these impact on approximations in  the physical domain. Most notably, we establish that singularities away from the physical region control the radius of convergence for all field expansions.

The outline of the paper is as follows. In Sec.~\ref{sec:model} we briefly introduce the effective action of 
$O(N)$-symmetric scalar theories and discuss the RG flows and the classical fixed points. In 
Sec.~\ref{sec:GBM}, we discuss fluctuation-induced fixed points due to the longitudinal Goldstone modes including the Wilson-Fisher fixed point, tri-critical fixed points, and Gaussian fixed points.
We provide exact recursion relations for couplings and consider both global analytical solutions as well as local analytical expansions about small or large fields.  We repeat the analysis for the case of transversal radial fluctuations of the fields in  Sec.~\ref{sec:radial mode}, and compare with the previous findings. 
We conclude in Sec.~\ref{Conclusion} and defer some technicalities to an appendix~\ref{AppA}.

\section{\bf Renormalisation group}
\label{sec:model}
In this section we recall the basic set-up and introduce the main equations and approximations.

\subsection{Functional renormalisation}
Functional renormalisation is based on a Wilsonian version of the path integral where parts of the fluctuations have been integrated out  \cite{Wilson:1973jj,Polchinski:1983gv,Wetterich:1992yh,Ellwanger:1993mw,Morris:1993qb,Litim:2000ci}. More concretely, we consider Euclidean scalar field theories with partition function \cite{Wetterich:1992yh}
\begin{equation}
Z_k[J]=\int D\varphi\exp(-S[\varphi]-\Delta S_k[\varphi]-\varphi\cdot J)\,.
\end{equation}
Here $S$ denotes the classical action and $J$ an external current. The expression differs from a text-book partition function $Z[J]$ through the presence of the Wilsonian cutoff term  
\beq
\Delta S_k[\varphi]=\frac12\int\,\frac{d^dq}{(2\pi)^d}\,\varphi(-q)\, R_k(q^2)\,\varphi(q)
\eeq
in the action \cite{Wetterich:1992yh}. The function $R_k(q^2)$ is chosen such that low momentum modes $q^2\ll k^2$ are suppressed in the path integral, but high momentum modes $q^2\gg k^2$ propagate freely \cite{Berges:2000ew}.
For our purposes, we require $R_k(q^2\to 0)>0$ for $q^2/k^2\to 0$ and $R_k(q^2)\to 0$ for $k^2/q^2\to 0$ to ensure that $R_k$ acts as an IR momentum cutoff which is removed in the physical theory $(k\to 0)$ \cite{Litim:2001up,Litim:2000ci,Litim:2001fd}.

It is convenient to replace the partition function by the `flowing' effective action $\Gamma_k$ to which it relates via a Legendre transformation $\Gamma_k[\phi]=\sup_J(-\ln Z_k[J]+\phi \cdot J) +\Delta S_k[\phi]$, where $\phi=\langle\varphi\rangle_J$ denotes the expectation value of the quantum field. The scale-dependence of $\Gamma_k$ is given by an exact functional identity \cite{Wetterich:1992yh} (see also \cite{Ellwanger:1993mw,Morris:1993qb})
\begin{equation}\label{FRG}
\partial_t\Gamma_k=\frac12\Tr\frac{1}{\Gamma_k^{(2)}+R_k}\partial_t R_k\,,
\end{equation}
relating the change of scale for $\Gamma_k$ with an operator trace over the full propagator multiplied with the scale derivative of the cutoff itself.  The convergence and stability of the RG flow is controlled by the regulator $R_k$ \cite{Litim:2000ci,Litim:2001up}. Optimised choices are available and allow for analytic flows and an improved convergence of systematic approximations \cite{Litim:2001fd,Litim:2001dt,Litim:2010tt}.

The flow \eq{FRG} has a number of interesting properties. By construction, the flowing effective action interpolates between the classical  action for large RG scales and the full quantum effective action $\Gamma$ in the infrared limit $k\to 0$. At weak coupling, it reproduces the perturbative loop expansion \cite{Litim:2001ky,Litim:2002xm}. It relates to the well-known Wilson-Polchinski flow ~\cite{Polchinski:1983gv} by means of a Legendre transformation, and reduces to the Callan-Symanzik equation in the limit where $R_k(q^2)$ becomes a momentum-independent mass term \cite{Litim:1998nf}. The RHS of the flow \eq{FRG} is  local in field- and momentum space due to the regulator term, enhancing the stability of the flow \cite{Litim:2005us}. This also implies that the change of $\Gamma_k$ at scale $k$ is mainly governed by loop-momenta of the order of $k$.

\subsection{Derivative expansion}
We are  interested in critical $O(N)$ symmetric scalar field theories to leading order in the derivative expansion. The derivative expansion is expected to have good convergence properties because the anomalous dimension of scalar fields at criticality are of the order of a few percent. This expectation has been confirmed quantitatively based on studies up to fourth order in the expansion  \cite{Litim:2010tt}.
The main purpose of the present paper is to analyse the leading order in the derivative expansion analytically in view of global aspects of fixed points. The flowing effective action is approximated by
\begin{equation}
\Gamma_k=\int d^3x\left(
	\frac 12 \partial_\mu\vec \phi \, \partial_\mu \vec\phi + V_k(\phi)
		\right)\,,
\end{equation}		
where $\vec \phi$ is a vector of $N$ scalar fields. 
Using the optimised regulator proposed 
in \cite{Litim:2002cf}, the momentum trace is performed analytically.
We also introduce the dimensionless effective potential
$u(\rho)=V_k(\phi)/k^3$ and dimensionless fields $\rho=\frac 12 \phi^2/k$ which accounts for the invariance of the action under reflection in field space $\phi\rightarrow -\phi$. 
The RG flow of the potential is then given by the partial differential equation
\begin{equation}\label{flow}
\partial_t u =-3u+\rho u^\prime 
+(N-1)I[u']+I[u'+2\rho u'']\,,
\end{equation}
where $t=\ln k$ denotes the logarithmic RG `time'. 
The first two terms on the RHS arise due to the canonical dimension of the potential and of the fields, whereas the third and fourth term arise due to fluctuations. The functions $I[x]$ arise from the Wilsonian momentum cutoff and relate to the loop integral in \eq{flow}. Explicitly,
\begin{equation}\label{Igen}
I[x]=k^{-d}\int d^dq\, \frac{\partial_t R_k(q^2)}{q^2+R_k+x\, k^2}\,.\end{equation}
In the present calculation we are going to use the optimised regulator
\begin{equation}\label{opt}
R_{k}(q^2)=(k^2-q^2)\,\theta(k^2-q^2)\,,
\end{equation}
following \cite{Litim:2002cf,Litim:2001up,Litim:2000ci}. Then the integral \eq{Igen}  can be evaluated analytically, leading to 
\beq\label{I}
I[x]=A_d/(1+x)\,.
\eeq
Here, $A_d=2/(d\,L_d)$ and $L_d=(4\pi)^{d/2}\Gamma(d/2)$ denotes the $d$-dimensional loop factor arising from the angular integration  of the operator trace, with  $A_3=1/(6\pi^2)$.
The prefactor is irrelevant for all technical purposes. For convenience we scale it into the potential and the fields via $u\to u/A_d$ and $\rho\to\rho/A_d$, meaning $A_d\to 1$ in \eq{I}. The benefit of this normalisation is that couplings, at interacting fixed points, are now measured in units of the appropriate loop factors.

The flow is driven by the $N-1$ Goldstone modes and the radial mode, which are responsible for the fluctuation-induced terms on the RHS of \eq{flow}.
 The flow for the first derivative of the potential is given by
\begin{eqnarray}\label{eq:flowprime}
\partial_t u'  &=&-2u'+\rho u''
+(N-1)u'' I'[u']
+(3u''+2\rho\,u''')I'[u'+2\rho u'']\,.
\end{eqnarray}
In the remaining parts of the paper, we are interested in the Wilson-Fisher fixed point solutions $\partial_t u'=0$ of \eq{eq:flowprime} for all fields,  and for all universality classes $N$.
The fixed point potential obeys an ordinary second order non-linear differential equation given by \cite{Litim:2000ci,Litim:2001up,Litim:2002cf}
\begin{eqnarray}\label{utriple}
2\rho \frac{du''}{d\rho}&=&-\left[3-(N-1)\frac{(1+u'+2\rho u'')^2}{(1+u')^2}\right]\,u''
\label{dglFP}
+(2u'-\rho\,u'')(1+u'+2\rho\,u'')^2\,.
\end{eqnarray}
Eq.~\eq{dglFP}  determines the Wilson-Fisher fixed point, subject to the unique Wilson-Fisher boundary condition fixing, eg.~$\rho_0$ with $u'(\rho_0)=0$ and $u''(\rho_0)$.
In the infinite-$N$ limit, the fixed point equation becomes first order and reads
\begin{eqnarray}\label{udouble}
\frac{du'}{d\rho}&=&\frac{2u'(1+u')^2}{\rho(1+u')^2-1}
\label{dglFPN}
\end{eqnarray}
after a simple rescaling with $N$. The universal physics in these theories is characterised by scaling 
exponents $\vartheta_n$, defined as 
$\partial_t \delta u'_n=\vartheta_n\delta u'_n$ at scaling, 
where $\delta u'_n(\rho,u')$ are the eigensolutions at the fixed point. 
Determining $u'$ accurately is central in obtaining accurate estimates for 
the universal numbers $\vartheta_n$ \cite{Litim:2001dt,Litim:2003kf,Litim:2010tt}. 

\subsection{Classical fixed points}
Prior to a discussion of the fluctuation-induced fixed points of the theory we begin with the classical fixed points. In the absence of fluctuations we have  $I=0$,  and the RG flow \eq{eq:flowprime}  becomes
\beq\label{classical}
\partial_t u'+2u'-(d-2)\rho u'' =0
\eeq
in $d$ euclidean dimensions. This RG flow has a Gaussian fixed point
\beq\label{Gaussclassical}
u'_*\equiv 0
\eeq
in consequence of \eq{classical} being linear in $u'$. From  the RG flow of the inverse $1/u'$,
\beq\label{classicalinv}
\partial_t \01{u'}-2\,\0{1}{u'}-(d-2)\rho\,\partial_\rho\left( \01{ u'} \right)=0
\eeq
we conclude that the theory also displays an `infinite' Gaussian fixed point 
\cite{Nicoll:1974zza}
\beq\label{infGaussclassical}
1/u'_*\equiv 0\,.
\eeq
More generally, the RG flows \eq{classical} and \eq{classicalinv} are solved analytically by
\beq\label{solclassical}
u'=\rho^{2/(d-2)}K\left(\rho\, e^{(d-2)t}\right)
\eeq
for arbitrary functions $K(x)$ which is fixed only by the boundary conditions at $t=0$. 
Fixed point solutions are those which no longer depend of the RG `time' $t$,
and a trivial one is given by $K(x)=$~const. This leads to a line of fixed points
\beq\label{marginalclassical}
u'_*=c\,\rho^{2/(d-2)}
\eeq
parametrized by the value of the coupling $c$. We note that the canonical mass dimension of the coupling $c$ in $d$ dimensions vanishes, $[c]=0$. The linearity of the RG flow allows us to study the flow of field monomials independently, $u'=\lambda_n\rho^n$ (no sum). We find that
\beq
\lambda_n(t)=\lambda(0)\,e^{\vartheta_n\,t}
\eeq
where the eigenvalues are given by
\beq\label{thetaclassical}
\vartheta_n=-2+(d-2)n\,,
\eeq
which are the well-known classical eigenvalues at the (infinite) Gaussian fixed point. For $n>n_+=\frac{2}{d-2}$ $(n<n_+)$ the eigenvalue \eq{thetaclassical} is positive (negative), the corresponding coupling  IR repulsive (attractive). Furthermore, the IR attractive (repulsive) couplings approach the Gaussian (infinite Gaussian) fixed point in the IR limit. In this light the case $n=n_+$ \eq{marginalclassical} is marginal leading to a finite fixed point $u'_*$.

\section{\bf  Infinite $N$}\label{sec:GBM}

The purpose of this section is to include fluctuations, and to obtain and compare the full closed form for the three dimensional Wilson-Fisher fixed point solution in the large-$N$ limit with approximate analytical solutions based on various expansions in field space.

\subsection{Analytical fixed points}
\label{sec:Analytical solution}

\begin{figure}[t]
\includegraphics[width=.6\hsize]{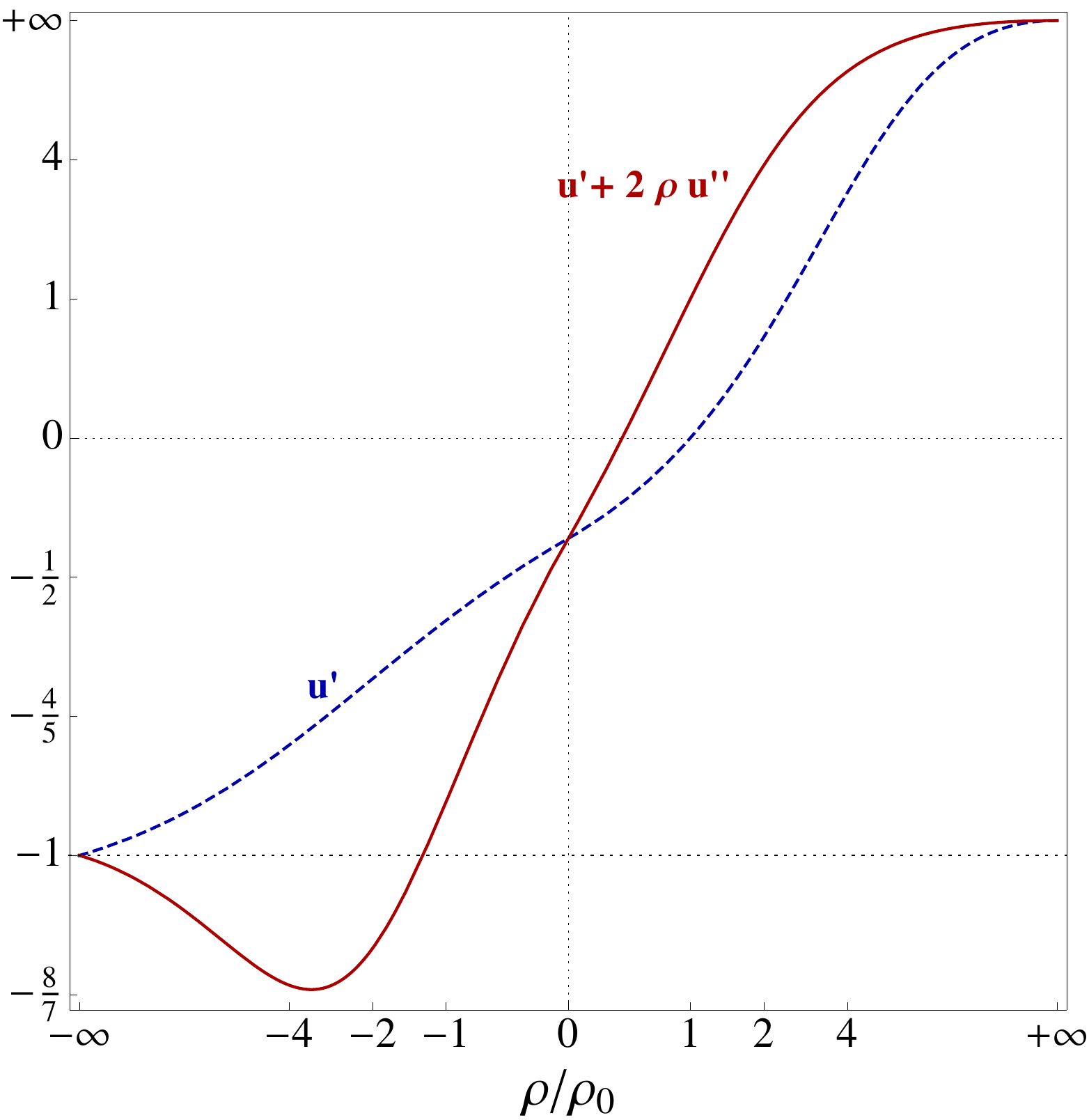}
\caption{The Wilson-Fisher fixed point $u_*'(\rho)$ in the infinite-$N$ limit (dashed line) and the amplitude $u'_*+2\rho u''_*$ (full line) for all values of the field squared.}
\label{Amplitude}
\end{figure}

We begin with the global fixed point solution in the limit where the transversal (Goldstone) 
modes dominate, corresponding to the 
massless excitations in the vicinity of the potential 
minimum. The longitudinal or radial mode, corresponding to the massive excitation 
at the potential minimum, is neglected and
will be considered in Sec.~\ref{sec:radial mode}.
In this limit, the local potential approximation becomes exact 
and the anomalous dimension of the field vanishes identically.
Formally, this approximation is achieved in the limit of the 
number of scalar fields $N\to\infty$, 
where the flow equation simplifies as
\begin{equation}
\label{flowLN}
\partial_t u=
-3u+ \rho u' +\frac{N}{1+u'}\,.
\end{equation}
We rescale the factor $N$ into the potential and the
fields $u\to u/N$ and $\rho\to \rho/N$ with $u'$ unchanged,
formally equivalent to setting $N=1$ in \eq{flowLN}. The flow for $u'$ then
reads
\begin{equation}
\label{flow'LN}
\partial_t u'= -2u'+ \rho u'' -\frac{u''}{(1+u')^2}\,.
\end{equation}
The flow equation~\eq{flow'LN} can be integrated analytically in closed form using the method of characteristics \cite{Litim:1995ex,Tetradis:1995br}. 
For $u'\ge 0$ the solution is
\begin{equation}\label{ExactLargeN}
\frac{\rho-1}{\sqrt{u'}}
-\frac{1}{2}\frac{\sqrt{u'}}{1+u'}
-\frac{3}{2}\arctan\sqrt{u'}=c\,,
\end{equation}
and the solution for $u'\le0$ follows from analytical continuation as
\begin{equation}\label{ExactLargeNnegative}
\frac{\rho-1}{\sqrt{-u'}}
+\frac{1}{2}\frac{\sqrt{-u'}}{1+u'}
-\frac{3}{4}\ln\frac{1-\sqrt{-u'}}{1+\sqrt{-u'}}=c\,.
\end{equation}
The coefficient $c$ is a free parameter. The solution with $c=0$ corresponds to the Wilson-Fisher fixed point.  It extends over 
all fields $\rho\in [-\infty,\infty]$, also exhausting the range of available values for $u'\in[-1,\infty]$ (see Fig.~\ref{Amplitude}). It leads to the universal scaling exponents
\beq
\label{spherical}
\vartheta=-1, 1, 3, 5, 7, \cdots
\eeq
corresponding to one critical direction with critical index $\nu=1$. The fixed point for $1/c\to 0$ is the  Gaussian fixed point \eq{Gaussclassical}, with Gaussian scaling exponents \eq{thetaclassical},
\beq
\label{Gauss}
\vartheta=-2, -1, 0, 1, 2, \cdots
\eeq
The two negative eigenvalues $-2$ and $-1$ relate to the mass term and the quartic coupling, and the zero eigenvalue relates to the exactly marginal $\phi^6$ coupling.   We mention for completeness that for finite $|c|>c_{\rm crit}$, some fixed point solutions correspond to a family of tricritical ones which equally display Gaussian scaling. These will not be discussed any further in this paper.
For the potential the additional eigenvalue $-3$ appears which is irrelevant in the absence of (quantum) gravity effects as it relates to overall shifts of the vacuum energy. 

\subsection{Algebraic fixed points}
We now discuss in more detail an iterative construction of fixed point solutions   which we shall use below \cite{Litim:2002cf}. We assume that the partial differential equation \eq{flow} can be transformed into a set of infinitely many coupled ordinary differential equations for suitable couplings $\{\lambda_n\}$. The specifications for this are irrelevant to the main argument, and explicit examples will be provided below. The RG flow is then determined by the RG flow for all couplings, the $\beta$-functions.  In terms of these, a fixed point has to obey
\begin{equation}\label{dlambdan}
\partial_t \lambda_n\equiv \beta_n(\{\lambda_i\})=0\,,
\end{equation}
for all $n$. In general, the $\beta$-functions \eq{dlambdan} depend on the couplings themselves. Provided that the $\beta$-function for any given coupling $\lambda_n$ depends only on finitely many coupling $\lambda_i$, \eq{dlambdan}  can be solved. This provides us with algebraic relations amongst the fixed point couplings of the form
\begin{equation}\label{lambda*}
\lambda_{n,*}= \lambda_{n,*}(\{\lambda_{i,*}\})\,.
\end{equation}
These relations amongst couplings must be fullfilled for any fixed point. These relations  can be further reduced upon iteration with $n$ \cite{Litim:2002cf}. This bootstrap-type strategy then provides us with expressions for (most of) the couplings $\lambda_n$ in terms of a few free parameters $c_\ell$ \cite{Litim:2002cf},
\begin{equation}\label{lambdac}
\lambda_{n,*}= \lambda_{n,*}(\{c_\ell\})\,.
\end{equation}
The free parameters $c_\ell$ are those couplings which remain undetermined by the iterative strategy, meaning that none of these are fixed by solving \eq{dlambdan} for any $n$. If so, \eq{lambdac} supplies us with a family of fixed point candidates, parametrised by the set of $L$ parameters $\{c_\ell,\ell=1,\cdots,L\}$. If $L>0$, it then remains to fix the remaining free parameters by other means to identify unique physical solutions.  By construction, the local couplings of any fixed point of the theory have to obey  \eq{lambda*}. We stress that this technique is not bound to critical scalar theories. For a recent applications to quantum gravity, see \cite{Falls:2013bv,Falls:2014tra}.

\subsection{Minimum}

Next we turn to systematic local expansions of the RG, starting with the expansion about the potential minimum denoted as expansion $A$. 
The existence of a non-trivial minimum in the fixed point solution follows from the RG flow at $u'=0$.  Therefore we can write the polynomial expansion as
\begin{equation}
\label{minimumlargeN}
u(\rho)=\sum_{n=2}^\infty~\frac{\lambda_n}{n!}~(\rho-\vev)^n\,.
\end{equation}
Subsequently we truncate \eq{minimumlargeN} at some maximum order in the expansion. Prior to discussing the solutions, it is interesting to consider the $\beta$-functions for the relevant and marginal couplings of the ansatz \eq{minimumlargeN}. Using the RG flow, we have
\begin{eqnarray}
\partial_t\rho_0&=&1-\rho_0\nonumber\\
\partial_t\lambda&=&-\lambda(1-2\lambda)\label{dlambda}\\
\partial_t \tau&=&-6\lambda(\lambda^2-\tau)\nonumber
\end{eqnarray}
where we used $\lambda\equiv \lambda_2$ and $\tau\equiv\lambda_3$. Note that the RG flow of the vacuum expectation value (VEV) $\rho_0$  and for the quartic interaction fully decouple from the system. The flow for the VEV displays an IR repulsive fixed point at 
\beq\label{r*}
\rho_{0,*}=1\,.
\eeq
It also displays  two IR attractive fixed points at $1/|\rho_0|=0$, corresponding to the symmetric and the symmetry broken phases of the theory. The flow for the quartic coupling displays an IR repulsive fixed point at $\lambda=0$ and an IR attractive fixed point at 
\beq\label{la*}
\lambda_*=\frac12\,.
\eeq 
The latter is the Wilson-Fisher fixed point together with \eq{r*}, while the former corresponds to tricritical fixed points including the Bardeen-Moshe-Bender phenomenon. Finally, the RG flow of the sextic coupling is fully controlled by the quartic interactions. At the tricritical fixed point with \eq{r*} and $\lambda_*=0$, the sextic coupling becomes exactly marginal $\partial_t\tau\equiv 0$ leaving $\tau$ as a free parameter of the theory. On the other hand, at the Wilson-Fisher fixed point, the sextic coupling achieves the IR attractive fixed point 
\beq
\tau_*=\frac14\,.
\eeq 
This pattern is at the root for the entire fixed point structure of the theory. 

At either of the above fixed points, the expansion \eq{minimumlargeN} allows for a recursive solution of the fixed point condition for the flow \eq{flow'LN} in terms of the polynomial couplings. 
At the tricritical point, solving \eq{dlambdan} recursively, all higher couplings $\lambda_n$ with $n>3$ become functions of the exactly marginal coupling $\tau$. At the Wilson-Fisher fixed point, remarkably, the recursive fixed point solution is unique to all orders and free of any parameters. This is a consequence of the decoupling of both the VEV and the quartic interactions. Specifically, using the RG flow  and the expansion \eq{minimumlargeN}, the general recursive relation for the polynomial couplings $\lambda_n$  (for $n\ge3$) is given by 
\begin{eqnarray}
\lambda_n &=& \frac{1}{3-2n} \bigg[
\frac{n}{2}(n-4) \lambda_{n-1}
+ \sum_{k=2}^{n-2} \binom{n}{k}
\lambda_{k+1}
[\lambda_{n-k+1} + (n-k-3)\lambda_{n-k}]
\bigg]\,,
\end{eqnarray}
together with $\vev=1$ and  $\lambda_2=\frac12$. These expressions are straightforwardly generalised to $d\neq 3$ dimensions. Explicitly, for the first few couplings at the Wilson-Fisher fixed point, we find
\begin{eqnarray}\label{lambdaA}
\vev&=&1\,,\quad\lambda_2=\frac12\,,\quad\lambda_3=\frac14\,,\quad \lambda_4=\frac3{40}\,,\quad
\lambda_5=-\frac3{112}\,,\quad\lambda_6=-\frac{29}{1120}\,,
\end{eqnarray}
and similarly to higher orders. 
Furthermore, the universal scaling exponents, the eigenvalues of the stability matrix
\begin{equation}\label{M}
M_{ij}=\left.\frac{\partial \beta_i}{\partial\lambda_j}\right|_*
\end{equation}
come out exact at each and every order in the polynomial approximation,
\beq
\vartheta=-1,1,3,5,7,\cdots
\eeq
and agree, as they must, with those of the spherical model. 

It is noteworthy that the recursive solution provides us with an exact, parameter-free fixed point solution for the couplings \eq{lambdac} within the small-field regime. Its domain of validity is limited due to a finite radius of convergence of the expansion: no free undetermined parameters remain.  Empirically, the absolute values of the polynomial couplings \eq{lambdaA} grow, roughly, as
 \beq\label{growth}
|\lambda_n|\approx \frac{(n-1)!}{\pi^n \ln(2 \pi n)}
\eeq
for large $n$, suggesting that the radius of convergence $R_A$ is close to $\pi$.  The sign pattern of couplings is close  to $(++--)$. Small deviations from this pattern allow an accurate estimate for the location of a convergence-limiting singularity in the complex field plane.   In fact, the sign pattern, to very good accuracy, it is given by 
\begin{equation}\label{signpattern}
{\rm sgn}(\lambda_n)\approx \cos(n \phi_0 -\phi_1)\,,
\end{equation} 
where $(\phi_0,\phi_1)\approx(\frac{17}{31}\pi,19\pi)$.We therefore expect  that the convergence-limiting singularity in the complex plane is close to the imaginary axis, under the angle close to $\phi_0\approx 98.71^\circ$ from the expansion point $\rho_0$. 

\begin{figure}[t]
\begin{center}
\includegraphics[width=.62\hsize]{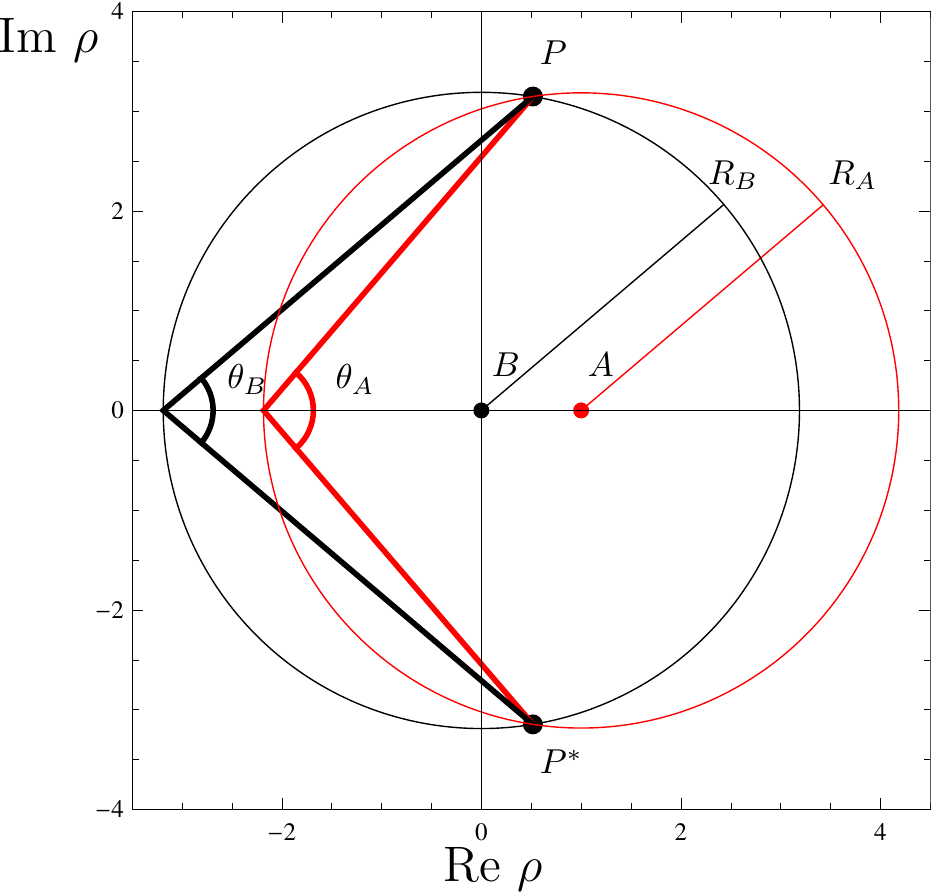}
\vskip.3cm
\caption{Singularity of the Wilson-Fisher fixed point in the complex field plane for small fields.  Shown are the radii of convergence for the expansions $A$ and $B$ of the Wilson-Fisher fixed point solution, and  the location of the convergence-limiting poles $P$ and $P^*$ (dots) in the complexified $\rho$-plane. Within errors, both expansions point towards the same singularity in the complex plane. }
\label{pZero}
\end{center}
\end{figure}

For a numerical determination of the radius of convergence, we note that the standard criteria for convergence such as root or ratio tests are not applicable. For a precise determination of the radius of convergence we therefore adopt a
criterion by Mercer and Roberts \cite{Mercer:1990}  detailed in App.~\ref{AppA}.  The criterion is designed for series which are governed by a pair of complex conjugate singularities. It offers estimates for the radius of convergence $R$, the angle $\theta$ under which the convergence-limiting pole occurs in the complexified $\rho$-plane, and the nature of the singularity $\nu$. When applied to the problem at hand, and based on the first 500 coefficients of the expansion, we find
\bea
R_A&=& 3.1835(2)\nonumber \\
\theta_A&=& 98.74(006)^\circ \label{radiusA} \\
 \nu_A&=&0.50(7)\,.\nonumber
\eea
The error estimate arises from varying the number of coefficients retained for the numerical fit. We conclude that the polynomial expansion about the potential minimum determines the fixed point solution exactly in the entire domain
\beq\label{domainA}
1-R_A\le\frac{\rho}{\rho_0}\le 1+R_A\,.
\eeq
The radius of convergence is limited through a square-root type singularity in the complex plane, whose location is approximately given by \eq{radiusA}. Graphically, this result is displayed in Fig.~\ref{pZero}. We will further exploit this result below to connect the fixed point solution in the small-field region to its large-field solution.

\subsection{Vanishing field}
\label{sec:Ninfty_vanishingfield}

For sufficiently small fields $|\rho|$, the flow \eq{flow'LN} admits a Taylor expansion in powers of the fields $\sim \rho^n$ to which we refer as expansion $B$.
We write our Ansatz for $u'$ as
\begin{equation}
\label{ZeroLargeN}
u(\rho)=\sum_{n=0}^\infty ~ \frac{\lambda_n}{n!} ~ \rho^n\,.
\end{equation}
Approximating the action up to the order $M$, the fixed point condition $\partial_t u'=0$ translates into $M+1$ equations $\partial_t\lambda_n=0$. The flow for the potential minimum $\lambda_0$ is irrelevant because the physics is invariant under $\lambda_0\to\lambda_0+c$. Following  \cite{Litim:2002cf}, and adopting the same  reasoning as previously, the algebraic equations for the couplings $\lambda_n$ are solved recursively in terms of the mass parameter at vanishing field,
\begin{equation}\label{mass}
u'(0)= \lambda_1  \equiv \mass\,.
\end{equation} 
The reason for the appearance of an  undetermined parameter is twofold. Firstly, the RG flow of none of the local couplings at vanishing field decouples from the remaining couplings -- in contradistinction to the RG flow of couplings at the potential minimum, see \eq{dlambda}. Secondly, the recursive solution is  simplified because the fixed point equation, at vanishing field, is effectively one order lower in derivatives. Consequently, the recursive solution retains one (rather than two) free parameters. Specifically, the couplings $\lambda_{n+1}$ for $n\ge 1$ are determined recursively from the lower-order couplings $\lambda_i$ with $i\le n$  and the  parameter \eq{mass} as
\beq\label{recursive}
\lambda_{n+1} = (1+\lambda_1) \bigg[ (n-3)\lambda_n
+ \sum_{k=0}^{n-1} \binom{n}{k} (n-k-3)\lambda_{k+1}\lambda_{n-k} \bigg] \, .
\eeq
We stress that similar recursion relations hold true for dimensions different from $d = 3$, for different regulator functions, and away from fixed points where $\partial_t u’ \ne 0$. 
Solving \eq{recursive} from order to order, we obtain the explicit expressions \cite{Litim:2002cf}
\begin{eqnarray}\label{coupNlarge0}
\lambda_0 &=& \ \ \s0{1}{3} (1+\mass)^{-1}\, \nonumber \\
\lambda_2 &=& -2 \mass (1+\mass)^2\, \nonumber\\ 
\lambda_3 &=&\ \  \, 2 \mass(1+\mass)^3 (1+5\mass)\,\nonumber \\
\lambda_4 &=& - 24 m^4 (1+\mass)^4 (1+3\mass) \label{lambdaB} \\
\lambda_5 &=&\ \  \,  48 m^6 (1+\mass)^5 (5+13\mass)\,,\nonumber \\
\lambda_6 &=& -48 m^6 (1+\mass)^6 (-5+34 \mass+119 m^4 ) \nonumber 
\end{eqnarray}
for the first few coefficients, and similarly to higher order. Structurally, the  solution has the form
\begin{equation}\label{lambdam}
\lambda_n= m^2(1+m^2)^n\,P_n(m^2)
\end{equation}
for all $n\ge 2$. Here the $P_n(m^2)$ are polynomials of degree $(n-2)$ in $m^2$. Using the notation $[n/2]=n/2$ $(n/2+1/2)$ for even (odd) $n$, we find that $P_n$ also contains an additional factor $(m^2)^{[n/2]-1}$ times a remaining polynomial $Q_n$
which has no further zeros at $m^2=0$ or at $m^2=-1$. 
There is a unique choice for $m^2$ corresponding to the Wilson-Fisher solution. Furthermore, there is a range of values for $m^2$ which corresponds to the tri-critical fixed points. We also recover the (trivial) Gaussian fixed point
\beq\label{GaussB}
m^2=0\,,
\eeq
which enforces the vanishing of all higher order couplings. In either of these cases the global solution extends over all fields. 
Interestingly, the series expansion also displays the exact non-perturbative fixed point
\beq \label{convexInf}
m^2=-1
\eeq
which entails the vanishing of all higher order couplings. It is responsible for the approach to convexity in a phase with spontaneous symmetry breaking \cite{Tetradis:1992qt,Litim:2006nn}. The convexity fixed point is only visible in the inner part of the effective potential $\rho<\rho_0$ and as such cannot extend over the entire field space. For the same reason the convexity fixed point is not visible in the expansion about the potential minimum discussed in the previous section.

 \begin{figure}[t]
\begin{center}
\includegraphics[width=.6\hsize]{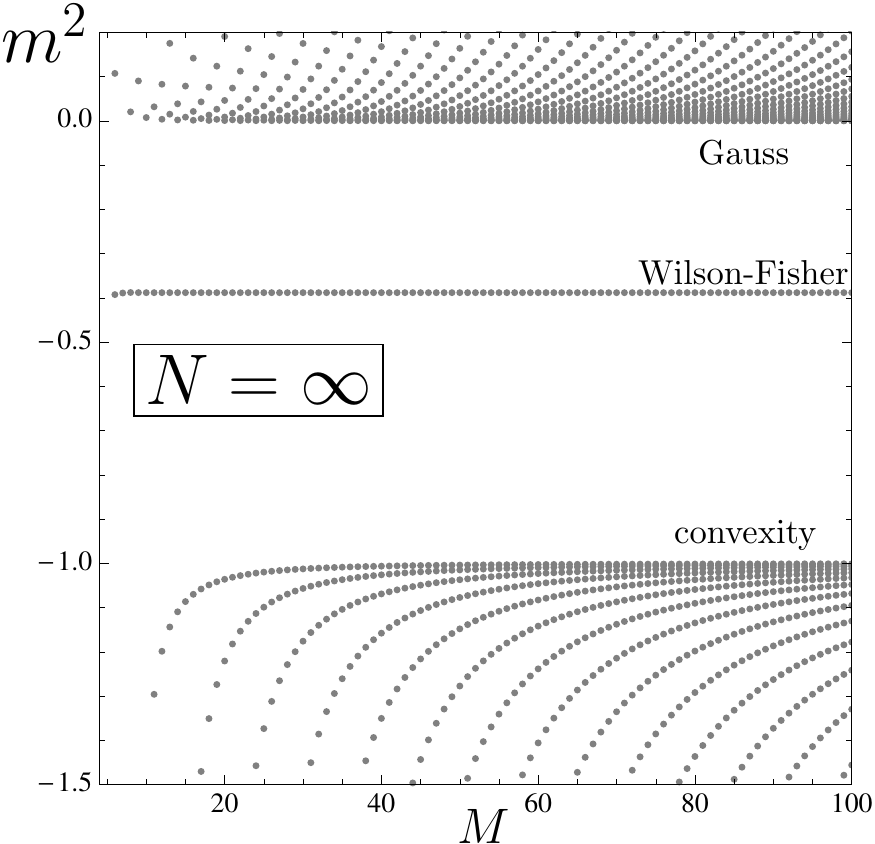}
\vskip.3cm
\caption{Shown are all real roots $m^2$ of the auxiliary condition \eq{auxInf} different from the exact roots at $0$ and $-1$, and  as a function of the approximation order $M$, with fixed point couplings given by the expressions \eq{SmallFieldSolution}.   The values for admissible real roots accumulate  close to the  Gaussian \eq{GaussB} and  the convexity fixed point \eq{convexInf}. The Wilson-Fisher fixed point \eq{massB} appears as a unique isolated root to each and every order in $M$.}
\label{pRootsInf}
\end{center}
\end{figure}

Returning now to the Wilson-Fisher fixed point, it remains to determine the fixed point value for the mass parameter which remained undetermined by the algebraic solution. This can be done either by exploiting the global analytical solution, an auxiliary condition at vanishing field, or results from the expansion about the minimum. From the analytical solution, we find the mass term from solving a transcendental equation $H(m^2)=-1$, with
\begin{equation}\label{H}
H(x)= \frac{1}{2} \frac{x}{1+x}
+\frac{3}{4} \sqrt{-x} ~ \ln \frac{1-\sqrt{-x}}{1+\sqrt{-x}} \,.
\end{equation}
The unique solution reads 
\begin{equation}\label{massB}
\mass=-0.388\,346\,718\,912\,782\,\cdots\,.
\end{equation}
Alternatively we may exploit the expressions \eq{lambdam} and impose the auxiliary condition that \beq\label{auxInf}
\lambda_{M+1}(m^2)=0
\eeq
at approximation order $M$ to determine $m^2$. Fig.~\ref{pRootsInf} shows the roots of the first 100 couplings. In the figure, we  have suppressed the trivial multiple roots at \eq{GaussB} and \eq{convexInf} corresponding to the exact Gaussian and convexity fixed points, respectively. Interestingly, the Gaussian and the convexity fixed point still appear in the remaining spectrum of fixed point candidates as accumulation points. In turn, the Wilson-Fisher fixed point appears with a unique solution to each and every order. Numerically, the WF root converges very rapidly towards the value \eq{massB}.

 Finally, we can exploit that the point $\rho=0$ lies within the radius of convergence of the expansion $A$ \eq{domainA}, which allows us to determine $m^2$ from the known expansion coefficients \eq{lambdaA}. Doing so, we find that the rate of convergence is fast: From the first 300 couplings of the expansion \eq{lambdaA}, we checked that the first $n$ terms suffice to reproduce $n/2$ significant figures of the parameter  \eq{massB}.
  
Next we estimate the radius of convergence for the expansion $B$. Using \eq{massB} together with the first 500 terms of  \eq{lambdaB}, we find that the coefficients grow as in \eq{growth} for large $n$, suggesting that the radius of convergence is close to $R_B\approx \pi$.  The sign pattern  is again close to $(++--)$. This time, we find \eq{signpattern}  with $\phi_0\approx\frac{14}{31}\pi$ which is approximately $81^\circ$. We therefore expect  that the convergence-limiting pole in the complex plane is close to the imaginary axis. Using the Mercer-Roberts technique with the help of the first 500 terms of the expansion, a more accurate estimate is found to be
\bea
R_B&=&3.1886(0)\nonumber\\
\theta_B&=& 80.682(4)^\circ\label{radiusB} \\
\nu_B&=&0.50(8)\nonumber
\eea
Comparing with \eq{radiusA}  we find similar, but not identical, radii of convergence $R_A\approx R_B$. The nature of the singularity appears to be of a square-root type in either case. Interestingly, the estimated singularity in the complex plane derived from either of the expansion \eq{radiusA} and \eq{radiusB} are the same, as can be seen in Fig.~\ref{pZero}. This is a strong hint towards the existence of a square-root singularity in complex field space in the full un-approximated fixed point solution. We discuss this in detail in Sec.~\ref{Singular} below.

We conclude that the polynomial expansion about  vanishing field determines the fixed point solution exactly in the entire domain
\beq\label{domainB}
-R_B\le\frac{\rho}{\rho_0}\le R_B\,.
\eeq
The overlap between the expansions \eq{minimumlargeN} and \eq{ZeroLargeN} therefore allow a complete analytical determination of the fixed point solution in the junction of \eq{domainA} and \eq{domainB}.

\subsection{Large fields}\label{sec:Ninfty_realfields}

For large fields $\rho/\rho_0\gg 1$, the effect of fluctuations is parametrically reduced
 and the effective potential approaches an infinite Gaussian fixed point  \cite{Nicoll:1975yq,Litim:2002cf,Litim:2005us}
\beq\label{infiniteGauss}
u'_*(\rho)=\gamma\,\rho^2\,,\quad\quad\rho/\rho_0\gg 1\,,
\eeq
where $\gamma$ is a free parameter, see \eq{marginalclassical}. The fixed point \eq{infiniteGauss} is  Gaussian in the strict sense that the contributions of fluctuations to the RG flow are parametrically  switched off. The fixed point scaling is then a consequence of the canonical dimension of the fields. The infinite Gaussian fixed point \eq{infiniteGauss} is also approached from the Wilson-Fisher fixed point solution in the limit where $\rho_0/\rho\to 0$. Therefore we may recover the Wilson-Fisher fixed point by expanding the RG flow  about \eq{infiniteGauss} in terms of a Laurent series in inverse powers of the fields to which we will refer as expansion $C$. We write
\begin{equation}\label{LargeFieldsLargeN}
u'(\rho) = \gamma\,\rho^2\left[1+\sum_{n=1}^{\infty} ~ \gamma_n ~ \rho^{-n}\right]\,.
\end{equation} 
Inserting \eq{LargeFieldsLargeN} into \eq{flow'LN} with $\partial_t u'=0$ leads to equations $\partial_t\gamma_n$, which are solved recursively for fixed points. Alternatively this can be achieved by introducing
\beq\label{factor}
v(x)=u'(\rho)/\rho^2\,,\quad x=1/\rho
\eeq
and Taylor-expanding $v(x)=\sum_{n=0}\gamma_n\,x^n$ for $x\ll 1$ (and $\gamma\equiv \gamma_0$). The interpretation of \eq{factor} is that the infinite Gaussian fixed point is factored out. The recursive solution determines all couplings uniquely as functions of the free parameter $\gamma$. We have computed the  first 500 coefficients in this expansion.
The first few non-vanishing coefficients read
\begin{eqnarray}
\gamma_{5} &=& - \frac{2}{5 \gamma^2},~\gamma_{7} = \frac{4}{7 \gamma^3}, ~
\gamma_{9} = - \frac{2}{3 \gamma^4}, \nonumber \\[1ex]
\gamma_{10} &=& - \frac{7}{25 \gamma^4}, ~
\gamma_{11} = \frac{8}{11\gamma^5}, ~
\gamma_{12} = \frac{36}{35 \gamma^5},
\label{lambdaC}
\end{eqnarray}
and similarly to higher order. Note that the first four coefficients vanish identically. 

One may wonder how the scaling exponents vary with the free parameter $\gamma$.  The stability matrix  for the flows $\partial_t\gamma_n$, that is \eq{M} with $\lambda_n$ replaced by $\gamma_n$, has no entries on its upper diagonal, and the dependence on the free parameter $\gamma$ only appears on the lower off-diagonal elements. The eigenvalues then reduce to the diagonal elements given by the canonical mass dimension of the couplings $\gamma_n$, which are
\beq\label{thetalargefields}
\vartheta=0,-1,-2,-3,-4,-5,\cdots
\eeq
independently of the finite value $\gamma\neq 0$. The marginal eigenvalue $\vartheta=0$ signals that $\gamma \rho^2$ is an exactly marginal operator  at the infinite Gaussian fixed point \eq{infiniteGauss}. The non-vanishing eigenvalues measure the canonical dimension of the field monomials $1/\rho^n$, and the negative sign states that this fixed point is UV attractive in all couplings except for the marginal one. Note that the eigenvalues \eq{thetalargefields} agree with those found from the classical theory \eq{thetaclassical} for inverse powers in the fields. We conclude that the asymptotic expansion at exactly $1/\rho = 0$ is sensitive only to the classical scaling of operators. 

Interestingly, the global Wilson-Fisher fixed point solution connects to this set of fixed points for a specific value of the parameter $\gamma$, despite the fact that the scaling properties seem different. In fact, the Wilson-Fisher fixed point
corresponds to
\begin{equation}\label{gammaWF}
\gamma=\frac{16}{9\pi^2}=0.180\,126\,548\,697\,489\,\cdots\,.
\end{equation}
This unique value can be computed either from the closed analytical solution
\eq{ExactLargeN}, or from matching to the expansion $A$ using \eq{lambdaA} in a regime where both radii of convergence overlap. 

\begin{figure}[t]
\begin{center}
\includegraphics[width=.62\hsize]{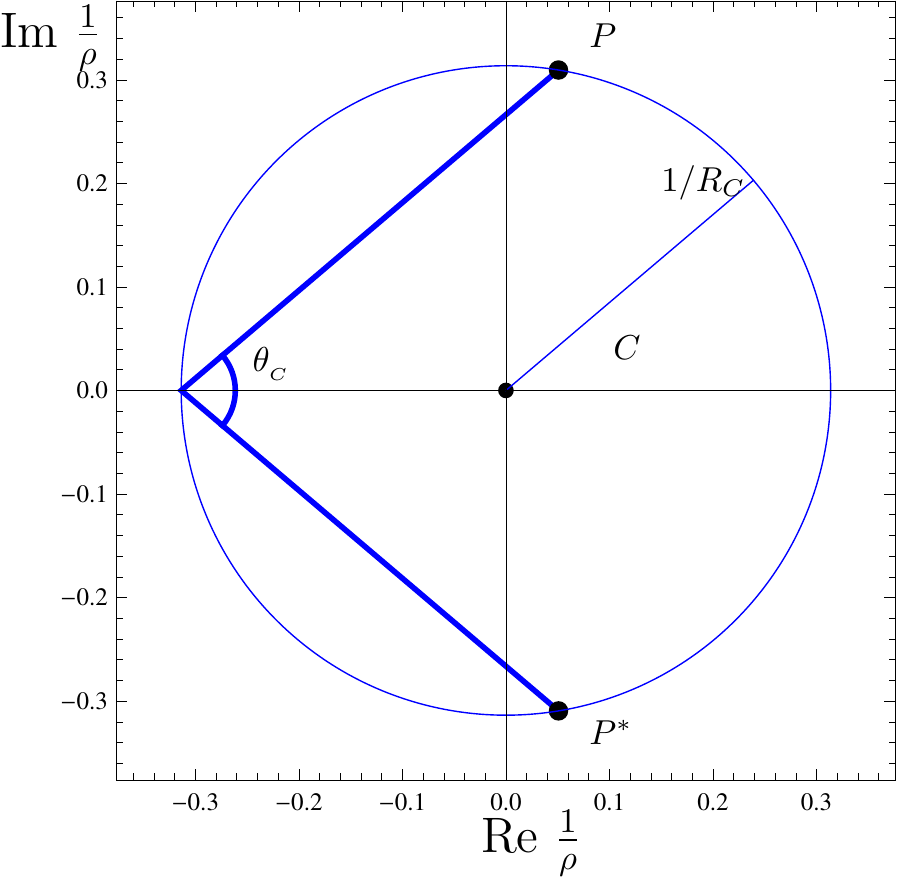}
\vskip.5cm
\caption{Singularity of the Wilson-Fisher fixed point in the complex field plane  for large fields. Shown are the radius of convergence for the expansion $C$  of the Wilson-Fisher fixed point solution and the estimated location of the convergence-limiting  pole (dots) in the complexified $\frac1\rho$-plane. Within errors, the location of the pole coincides with the singularity identified in Fig.~\ref{pZero} based on the expansions $A$ and $B$.} 
\label{pInf}
\end{center}
\end{figure}

Using the Mercer-Roberts technique as before, and employing \eq{gammaWF} as input, we estimate the radius of convergence of expansion C from the first 500 coefficients 
of the expansion \eq{LargeFieldsLargeN} as 
\bea
R_C&=&3.18(9)\nonumber\\
\label{radiusC}
\theta_C&=& 80.682(4)^\circ
\\ \nonumber
\nu_C&=&0.4(79)\,,
\eea
see Fig.~\ref{pInf}. Comparing the result \eq{radiusC}  with the expressions \eq{radiusA}  and \eq{radiusB}  we conclude that
\begin{equation}
R_B=R_C
\end{equation}
to within our numerical accuracy. In turn, $R_A$ and $R_B$ agree on the percent level but start differing at the permille level. Furthermore,  the angle and the radius under which the singularity appears for the Laurent series in $1/\rho$ and for the Taylor series about $\rho$ are identical, within our numerical accuracy, $\theta_B=\theta_C$. The accuracy for the nature of the singularity is smaller, presumably because the expansion point is far away from the singularity. Still, the result suggests that $\nu_B\approx \nu_C$.
We conclude that the Laurent series about asymptotic fields determines the fixed point solution exactly in the entire domain
\beq\label{domainC}
R_C\le\frac{\rho}{\rho_0}\,,
\eeq
and the overlap between the expansions \eq{minimumlargeN}, \eq{ZeroLargeN} and \eq{LargeFieldsLargeN} therefore allow a complete determination of the fixed point solution in the junction of \eq{domainA}, \eq{domainB} and \eq{domainC}.

\subsection{Singularities in the complex field plane}\label{Singular}

The results of the previous subsections made it clear that the global fixed point displays a singularity in the complexified field plane. If so, its properties can be deduced from the global solution. To that end, we write the Wilson-Fisher fixed point solution \eq{ExactLargeN}, \eq{ExactLargeNnegative} as
\begin{equation}\label{FP}
\rho=1+H(u')\,,
\end{equation}
where the function $H$ is given in \eq{H}.
One may as well adopt the alternative representation
\begin{equation}\label{H1}
H(u')=\frac{1}{2}\frac{u'}{1+u'}
+\frac{3}{2}\sqrt{u'}\arctan\sqrt{u'}\,.
\end{equation}
Both representations, connected by analytical continuation, display a pole at $u'=-1$, a branch cut for $u'<-1$, and have identical Taylor-expansions in $u'$. 

We allow both $\rho$ and $u'$, and thus $H$, to take complex values.  By virtue of \eq{FP}, we find that the pole at $u'=-1$ corresponds  to the limit $1/\rho=0^-$. As such it is infinitely far away from any finite expansion point on the positive real $\rho$-axis. We then find that the fixed point displays another singularity in the complex plane at finite $\rho=\rho_s$, where the complexified $u'\to u'_s=u'(\rho_s)$ remains finite, but both the real and the imaginary part of the complexified second derivative $u''\to u''_s$ diverge. From \eq{FP} we have that 
\begin{equation}\label{u''}
u''=1/H'(u')
\end{equation}
which can be converted into $u''(\rho)\equiv u''(u'(\rho))$ using the relation \eq{FP}. The condition for a singularity in the quartic self-interaction at $\rho_s$  
becomes
\begin{equation}\label{H'}
H'(u'_s)=0\,.
\end{equation}
It provides us with two equations for its real and imaginary part, respectively. Both of these equations are transcendental, and their solutions are obtained numerically with any desired precision. 
We find that \eq{H'} has a unique solution where the  first derivative of the potential $u'_s\equiv u'(\rho=\rho_s)$ takes the values
 \begin{equation}\label{u'c}
\begin{array}{rl}
{\rm  Re}\,u'_s&=-1.426\,842\,101\cdots
 \\[.5ex]
 {\rm Im}\,u'_s&=\pm0.362\,515\,957\cdots\,.
 \end{array}
 \end{equation}
 The two signs for the imaginary part reflect that the original differential equation is real, implying that singularities in the complex plane must exist in complex conjugate pairs. 
We also observe
 \begin{equation}\label{u'c2}
\begin{array}{rl}
|u'_s|&=\ \ \, \ \ \  1.472\,173\,971\cdots\\[.5ex]
 \arg u'_s&=\pm 165.744\,559\,508^\circ\,.
 \end{array}
 \end{equation}
For the location of the singularity \eq{u'c}  in field space, using \eq{FP},  we find
 \begin{equation}\label{rc}
\begin{array}{rl}
{\rm  Re}\,\rho_s&=\ \ \,0.516\,269\,206\cdots
 \\[.5ex]
 {\rm Im}\,\rho_s&=\pm 3.146\,581\,843\cdots
\,.
 \end{array}
 \end{equation}
The imaginary part of $\rho_s$ is very close to $\pi$. The comparatively smallness of the real part implies that the singularity in $u''$  occurs close to the imaginary field axis.

\begin{figure*}[t]
\begin{center}
{\includegraphics[width=.9\hsize]{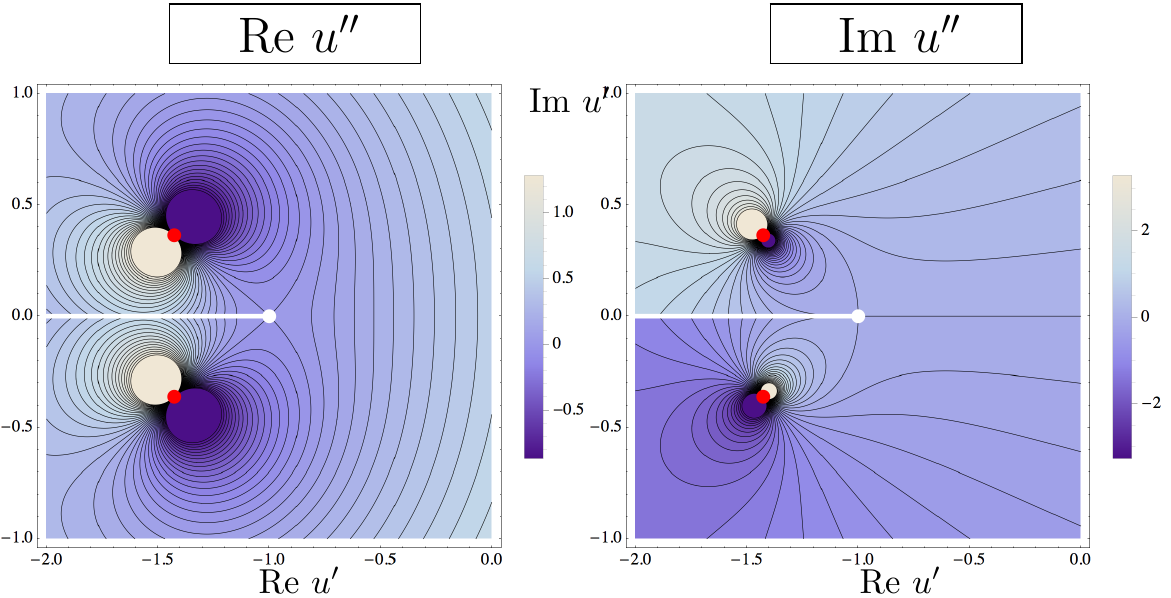}}
\vskip.3cm
\caption{Singularity of the Wilson-Fisher fixed point in the complex field plane. Shown are contour plots of the real (left panel) and imaginary part (right panel) of the field-dependent quartic self-coupling $u''$ as functions of the complexified field-dependent mass term $u'$, using \eq{H1}. The  full red dots
denote the location of the convergence-limiting singularities \eq{u'c} of  the global Wilson-Fisher fixed point solution. The horizontal white line indicates the cut at Re $u'<-1$ related to a discontinuity in Im~$u'$ across the cut. The full white dot indicates $u'=-1$.} 
\label{pSing}
\end{center}
\end{figure*}

We are now in a position to make the link with the results achieved previously based on the expansions $A, B$ and $C$. In the complex plane, the singularity $1/u''\to 0$ at \eq{rc} is the closest singularity to the expansion points $\rho=0$, $\rho=1$ and $1/\rho=0^+$, and thus its distance from the expansion point provides us with the radius of convergence. 
The  result \eq{rc}  translates into the exact radius of convergence $R_A=|\rho_0-\rho_s|$ and the exact angle $\theta_A=\arg(\rho_0-\rho_s)$ for the expansion $A$ about the potential minimum at $\rho_0=1$. Numerically, we have
  \begin{equation}\label{Aexact}
\begin{array}{rl}
R_A&=3.183\,547\,200
\\[.5ex]
\theta_A&=98.739\,781^\circ\,,
 \end{array}
 \end{equation}  
which is in excellent agreement with the estimate \eq{radiusA} derived from the recursive solution. 

Similarly, for the expansion $B$ about vanishing field $\rho_0=0$ we obtain the exact result $R_B=|\rho_s|$ and $\theta_B=\arg \rho_s$, meaning
\begin{equation}\label{Bexact}
\begin{array}{rl}
R_B&=3.188\,653\,507\\[.5ex]
\theta_B&=80.682\,326^\circ
 \end{array}
 \end{equation}
which compares very well with the estimate \eq{radiusB}. Notice that the expansion radii $R_A$ and $R_B$ are different, though only on the permille level, $R_B/R_A\approx 1.002$.

\begin{figure*}[t]
\begin{center}
\includegraphics[width=.9\hsize]{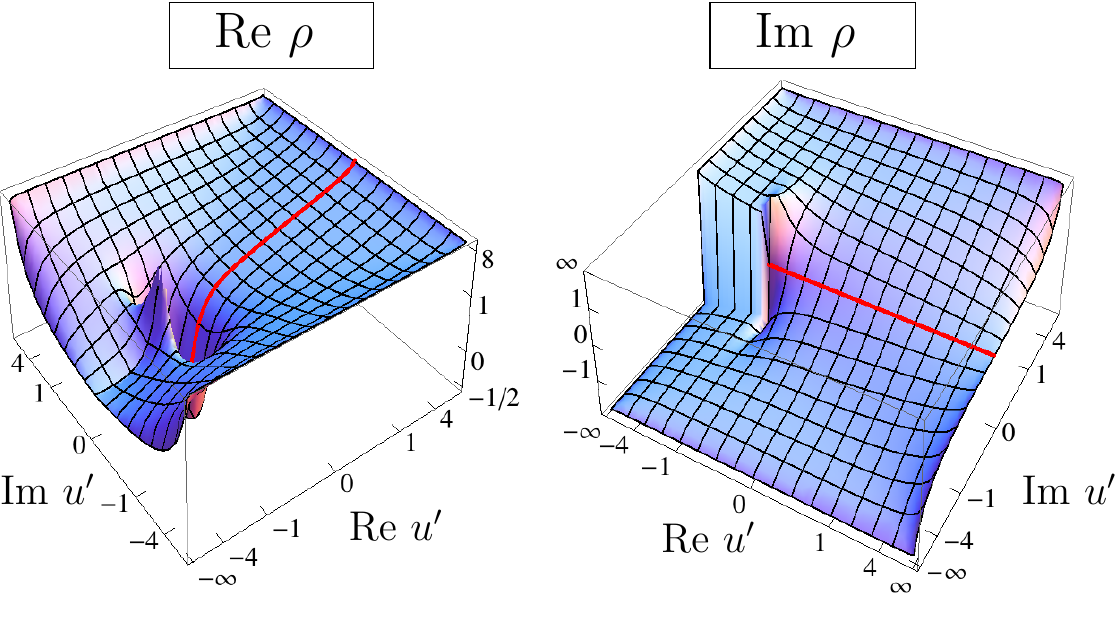}
\vskip.3cm
\caption{The global Wilson-Fisher fixed point in the entire complex field plane. Shown are the real (left panel) and imaginary (right panel) parts of the field $\rho$ as function of Re~$u'$ and Im~$u'$, including the pole at $u'=-1$. The full red line indicates the restriction of the fixed point solution to real  $-\infty\le \rho(u')\le \infty$.} 
\label{pRho}
\end{center}
\end{figure*}

For the expansion $C$ about asymptotically large field we obtain the radius from \eq{rc} as $1/R_C=1/|\rho_c|$. Thereby the complex conjugate poles exchange their places because $\arg (1/\rho_c)=-\arg\rho_c$ leaving the  angle $\theta_C=\theta_B$ unchanged, thus
\begin{equation}\label{Cexact}
\begin{array}{rl}
R_C&=3.188\,653\,507\\[.5ex]
\theta_C&=80.682\,326^\circ\,,
 \end{array}
 \end{equation}
in agreement with the estimate \eq{radiusC}. We also observe that $R_C=R_B$ exactly. It is quite remarkable that the  results for the various radii of convergence and loci of convergence-limiting poles agree very well with the analytical result derived from the global solution.

Finally, we determine the nature of the singularity $\nu$. To that end we expand the exact solution \eq{FP} in the vicinity of $\rho_s$ to find
\begin{equation}\label{vicinity}
\rho-\rho_s=\s012 H''_s\cdot (u'-u'_s)^2+{\rm subleading}\,,
\end{equation}
where $H''_s\equiv H''(u'_s)\neq 0$. Notice that a linear term in $(u'-u'_s)$ is absent at the singularity due to \eq{H'}. We conclude that
\begin{equation}\label{poleu'}
u'(\rho)-u'_s=\sqrt{\frac{2}{H''_s}}\,(\rho-\rho_s)^\nu
\end{equation}
close to the singularity, modulo subleading corrections of the order of $(\rho-\rho_s)^{2\nu}$. Hence the index  $\nu$ is found to be
\begin{equation}\label{nu12}
\nu=\frac12
\end{equation}
exactly, establishing that the singularity is of a square-root type. It controls a divergence which arises for the first time in the second derivative of the potential in the complex field plane
\begin{equation}\label{u2pole}
u''(\rho)=\frac{1}{\sqrt{2H''_s}}\frac{1}{\sqrt{\rho-\rho_s}}\,,
\end{equation}
and, subsequently, for all higher derivatives at $\rho_s$. The result \eq{nu12}  is in full agreement with the numerical estimates \eq{radiusA}, \eq{radiusB} and \eq{radiusC} for $\nu$ based on local expansion coefficients, establishing that the high-order polynomial expansion of the fixed point is reliable enough to also determine the nature of the singularity correctly. 

As an aside, we note that the radius of convergence $R'$ for an expansion of \eq{FP} in powers of the amplitude $u'$ about $u'=0$, given by the nearest pole in the complexified $u'$-plane, is controlled by the pole at $u'=-1$ rather than the one at \eq{u'c2}; hence $R'=1$.  

In Fig.~\ref{pSing}, we show contour plots of $u''$ expressed in terms of $u'$, \eq{u''}, and indicate the location of the pole and the reference points $u'=-1$ and $u'=0$. The real part of $u''$ is $Z_2$-symmetric under the map Im $u'\leftrightarrow-$Im $u'$. In this representation we find
\begin{equation}
u''=\frac{1}{H''_s}\,\frac{1}{u'-u'_s}
\end{equation}
from differentiating \eq{vicinity} with respect to the field. Notice that the square-root type singularity in $\rho$ \eq{u2pole} now becomes a simple pole in $u'$ as it must, following \eq{poleu'}.

\subsection{Imaginary fields}

Finally we turn to the regime of purely imaginary fields, corresponding  
to negative $\rho$. 
For small imaginary fields, the Taylor expansion in $\rho$ of 
Sec.~\ref{sec:Ninfty_vanishingfield} is applicable since the radius of 
convergence \eq{radiusB} is finite and extends to negative 
values. For large negative $\rho$  or large imaginary fields 
$\pm i\varphi\to \infty$, the effect of fluctuations is parametrically large. Furthermore, the presence of the fluctuation-induced  term $\sim u'' I'[u']$
in the  flow equation  implies that $-1\le u'<0$. This is different from the behaviour  
at large positive $\rho$ where $0<u'$ and the effects of fluctuations are suppressed, see Sec.~\ref{sec:Ninfty_realfields}.
We find that the fixed point  solution approaches 
\beq\label{convexity}
u'_*(\rho)\to -1
\eeq
for asymptotically large negative $\rho$,  thereby exhausting 
the domain of achievable values for $u'$.  Incidentally, this is also the non-perturbative fixed point of convexity, which is approached in a phase with spontaneous symmetry breaking. 

Since the Wilson-Fisher fixed point solution has $u'<0$ for field values below the VEV, it must be possible to re-cover it through an expansion about \eq{convexity},
which we denote as expansion $D$. The full asymptotic expansion of the Wilson-Fisher fixed point about \eq{convexity} for large negative $\rho$ contains inverse powers of the fields, powers of logarithms of the field, and products thereof. The set of non-trivial operators appearing in the fixed point solution is covered by the ansatz
\begin{equation}\label{AnsatzImaginaryLN}
u'(\rho)=-1+\sum_{m=1}^{\infty}\sum_{n=0}^{m-1}
\zeta_{m,n}\,\left(-\rho\right)^{-m} \,\ln^n(-\rho)\,.
\end{equation}
The structure of \eq{AnsatzImaginaryLN} can be understood as follows. In the limit where  $u'+1\to 0^+$, the first term in \eq{flow'LN} remains non-zero, while the second term is subleading. Therefore, the first and the third  term in  \eq{flow'LN}  have to cancel,
which is the case iff
\begin{equation}
0\le 1+u'=\frac{1}{2(-\rho)}+{\rm  subleading}\,.
\end{equation}
The next-to-leading term must contain a logarithm 
$\sim {\ln(-\rho)}/{(-\rho)^2}$ or 
else the fixed point condition cannot be satisfied. This leads to the pattern  
\eq{AnsatzImaginaryLN}. Adopting our strategy, we insert the Ansatz \eq{AnsatzImaginaryLN} up to order $n=M$ into \eq{ExactLargeNnegative} 
to find  $\frac{1}{2}M(M+1)$ algebraic equations for the expansion coefficients 
$\zeta_{m,n}$, all of which can be solved recursively in terms of a single free parameter \begin{equation}\label{zeta}
\zeta\equiv\zeta_{2,0}\,.
\end{equation} 
In terms of \eq{zeta}, the first few coefficients are explicitly given by
\begin{eqnarray}\label{zeta-LN}
\zeta_{1,0}&=&\ \  \, \012\,,\quad
\nonumber\\
\zeta_{2,1}&=&\ \  \, \038\,, \quad
\nonumber\\
\zeta_{3,0}&=&-\01{32}(9+32-64\,\zeta^2)\,,\nonumber\\
\zeta_{3,1}&=&\ \  \, \03{8}(4\,\zeta-1)\,,\quad
\nonumber\\
\zeta_{3,2}&=&\ \  \, \09{32}\,,
\label{zeta-LN}\\
\zeta_{4,0}&=&-\01{512} (99-400\,\zeta-2688\,\zeta^2+2048\,\zeta^3)\,,\quad\quad\quad
\nonumber\\
 \zeta_{4,1}&=&-\03{256}(25+336\,\zeta-384\,\zeta^2)\,,\nonumber\\
\nonumber
\zeta_{4,2}&=&\ \  \, \0{27}{256}(16\,\zeta-7)\,,
\\
\nonumber
\zeta_{4,3}&=&\ \  \, \0{27}{128}\,,\quad
\end{eqnarray}
and similarly to higher order. Using the exact result we establish
\begin{equation}
\zeta=\038(3\ln 2 -2)=0.029\,790\,578\,129\,938\,\cdots\,.
\end{equation}
To estimate the  radius of convergence for the expansion
$D$ we adopt an iterative version of the Mercer-Roberts technique to account for the logarithms.
We write  the series $u'=-1+\sum\tau_m(\rho)(-\rho)^{-m}$  in terms of coefficients $\tau_m(\rho)= \sum_{n=0}^{m-1} \zeta_{m,n} \ln^m(-\rho)$. Since the logarithm varies only slowly compared to powers, we approximate the coefficients $\tau_m=\tau_m(R^{(0)})$ for some trial coordinate  $\rho=R^{(0)}$ to determine the radius of convergence $R^{(1)}=f(R^{(0)})$, subject to the initial guess for the radius $R_D=R^{(0)}$. Subsequently the intial guess is  replaced by the first estimate $R_D=R^{(1)}$ to provide the input for the second estimate $R^{(2)}=f(R^{(1)})$ and so forth, until the procedure converges into a fixed point
\begin{equation}
R_D=f(R_D)\,.
\end{equation}   
Based on the first 100 coefficients $\zeta_{m,n}$ we find a rapid convergence with
\beq\label{radiusD}
R_D=-1.51(2)\,,
\eeq
implying that the expansion fully determines the solution in the domain
\beq\label{domainD}
\frac{\rho}{\rho_0}\le R_D\,.
\eeq
Most importantly, the domain overlaps with both small-field expansions $A$ given in \eq{domainA} and $B$, see \eq{domainB}. We therefore conclude that the unique Wilson-Fisher fixed point solution in the local expansion about the minimum $A$ actually fixes the entire fixed point solution globally, for all fields, with no further free parameter to be determined.

Fig.~\ref{pRho} shows the Wilson-Fisher fixed point solution for all complex fields and complex $u'$. The red line indicates the physically relevant part of the solution where $u'\ge -1$ and $\rho$ real. The pole at $u'=-1$ is clearly visible.  

\begin{figure}[t]
\includegraphics[width=.7\hsize]{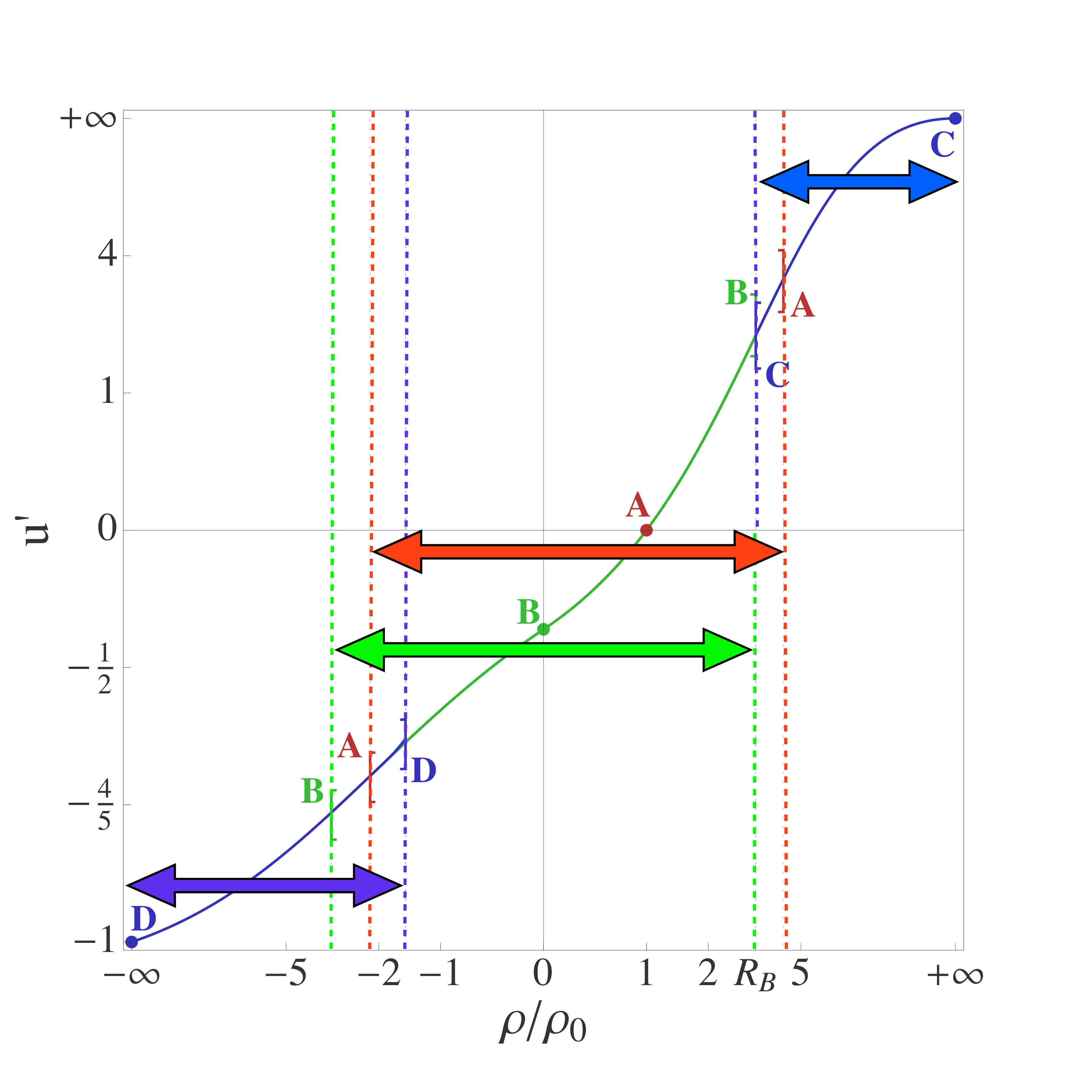}
\caption{Shown is the global Wilson-Fisher fixed point solution $u'_*(\rho)$  for real fields ($\rho\ge 0$) or purely imaginary  fields $(\rho<0)$, and the four local expansion regions  $A, B, C$ and $D$ together with their respective radii of convergence. The expansion $A$ yields a unique local Wilson-Fisher solution. Its overlap with $B, C$ and $D$  extends the local  to a global solution for all fields.}
\label{ErrorApprox}
\end{figure}

\subsection{Discussion}

The main results of this section are summarized in Fig.~\ref{ErrorApprox}, where we compare  the recursively-found fixed point solutions for the local expansions $A, B, C$ and $D$ of the Wilson-Fisher fixed point  with the global and analytically known solution $u_*'(\rho)$. The  local solutions \eq{minimumlargeN}, \eq{ZeroLargeN}, \eq{LargeFieldsLargeN} and \eq{AnsatzImaginaryLN} describe the scaling solution with high accuracy  within the respective domains of
applicability. The expansion points and the corresponding radii of convergence are also indicated  in Fig.~\ref{ErrorApprox}. Interestingly, the expansion $A$ about the local minimum leads to a  unique Wilson-Fisher fixed point solution, whereas the expansions $B, C$ and $D$ still depend on a free parameter $m^2, \gamma$ and $\xi$, respectively. However, since the radii of convergence of expansion $A$ overlaps with those of $B, C$ and $D$, the  parameters $m^2, \gamma$ and $\xi$ can be determined uniquely using only local information from the expansion $A$. In this sense, the local information about the Wilson-Fisher fixed point around  the local minimum of the potential suffices to fully determine the global fixed point solution. 

Furthermore, the universal scaling exponents are found reliably via the local small-field expansions $A$ and $B$, in agreement with \eq{spherical}. At asymptotically large fields, the global Wilson-Fisher solution asymptotes into a classical fixed point with classical exponents \eq{thetaclassical}. Strictly speaking, the  scaling behaviour of the spherical model is no longer visible at infinite field. This result is due to the fact that the functional RG flow \eq{FRG} is local both in momentum and in field space. Consequently, the non-trivial flow also becomes suppressed at asymptotically large fields. We conclude, therefore, that the universal scaling exponents are best deduced from the RG flow evaluated for fields of the size set by the RG scale parameter $k$. 

We also studied the  Wilson-Fisher fixed point solution for complex fields. Singularities in the complex field plane determine aspects of the fixed point solution for real field, in particular the radius of convergence of expansions in powers of the fields. Specifically, we identified
a square-root type singularity at the Wilson-Fisher fixed point in the complex $\rho$-plane, both within  the small and large field expansions, and from the closed analytical solution. Its location is fixed entirely through \eq{domainA}, \eq{domainB} and \eq{domainC}.  This result establishes that the convergence-limiting singularity in the complex plane can reliably be determined based on polynomial expansions in the field. In general, the validity of this observation may  depend on the choice for the regulator function. While it works very well for the optimised cutoff used here, it may fail for less suitable cutoffs such as eg.~sharp cutoff flows~\cite{Litim:2001up,Litim:2000ci,Litim:2001fd}.

\section{\bf Finite $N$}\label{sec:radial mode}

In this section, we study fluctuation-induced fixed points at finite $N$ by including corrections due to the longitudinal (or radial) fluctuations. Unlike at infinite $N$, a closed analytical solution to the RG flow is not available and we will instead resort to the recursive strategy adopted above for the expansions $A$, $B$, $C$ and $D$.\footnote{Notice that we will use a different normalisation of couplings at finite $N$, determined by the fixed point solution of \eq{eq:flowprime} with $A_d=1$ in \eq{I}. Effectively, this deviates from our convention for  infinite $N$ where the normalisation $A_d=1/N$ has been adopted; see \eq{flow'LN}. The reason for this is that we want to achieve finite expressions for couplings even if $N=0$ corresponding to the universality clafss of entfangled polymers. Whenever appropriate, we indicate how our findings in this section are related to those in the limit $N\to\infty$.}

\subsection{Minimum}

We begin with a polynomial expansion about the potential minimum
$\vev$ of the flow \eq{eq:flowprime}. This expansion is of the form
\begin{equation}\label{minifinin}
u (\rho) = \sum_{n=2}^{\infty} \frac{\lambda_n}{n!}~(\rho-\vev)^n\,.
\end{equation} 
and leads to coupled ordinary differential equations for the couplings and the VEV $\rho_0$. In particular, unlike the infinite-$N$ case \eq{dlambda}, the flow for the VEV no longer decouples,
\begin{eqnarray}
\label{dlambdaN}
\partial_t\rho_0&=&N-1-\rho_0+\frac{3 +2{\tau}\rho_0/{\lambda}}{(1+2\lambda\rho_0)^2}\,.
\end{eqnarray}
and similarly for the flow of the quartic $\lambda\equiv \lambda_2$ and the sextic coupling $\tau\equiv\lambda_3$. The explicit dependence of \eq{dlambdaN} on the quartic and sextic interactions implies that the recursive solution will at least depend on two free parameters.
By using the same iterative procedure for the Wilson-Fisher fixed point as before
we arrive at expressions for all fixed point couplings in terms exactly
two free parameters, $u''(\vev)\equiv \lambda$ and
the minimum $\vev$. In terms of these, higher order couplings $\lambda_i$ with $i>2$ are given recursively as
\begin{eqnarray}
\lambda_{n+2} =
\frac{1}{2\rho_0} 
\bigg[ 2(\rho_0\lambda_2 - n) \lambda_{n+1} 
+ (1+2\rho_0\lambda_2) \sum_{k=0}^{n-1} \sum_{l=0}^{n-k}\Omega_{n,k}\Upsilon_{n-k,l}\bigg]
\end{eqnarray}
where we used the abbreviations 
\begin{eqnarray}
\left. \begin{array}{rl}
\Omega_{i,j} =& \binom{i}{j}[\delta_j + (2j+1)\lambda_{j+1}+2\rho_0 \lambda_{j+2} ]
\\[2mm]
\Upsilon_{i,j} =& \binom{i}{j} [\delta_{i-j} + \lambda_{i-j+1}]
[\rho_0 \lambda_{j+1} + (j-3)\lambda_{j}] 
\end{array} \right.
\end{eqnarray}
and $\delta_j \equiv \delta_{j0}$. Resolving the recursive relations then leads to expressions of the form
\begin{equation}\label{lambdasNA}
\lambda_n =  \frac{\lambda}{(2\vev)^{n-2}}
~ Q_n(\lambda,\vev,N)
\end{equation} 
where $Q_n(\lambda,\vev,N)$ are $N$-dependent polynomials
in $\vev$ and $\lambda$. For example,
\begin{eqnarray}
Q_3&=&4\lambda\vev(1+\vev-N)(1+\vev\lambda) 
-(N+2)+\vev\,.
\end{eqnarray} 
Higher coefficients have a similar form but their expressions 
are too lengthy to be reproduced here. 
The unique Wilson-Fisher fixed point corresponds to a specific choice of parameters $(\lambda,\rho_0)$, for each $N$, which need to be determined by other means.

\subsection{Vanishing field}\label{finiteNvanish}

For small fields, the flow \eq{eq:flowprime} is
solved by Taylor-expanding the potential as
\begin{equation}\label{poly}
u(\rho)=\sum_{n=1}^{\infty}~\frac{\lambda_n}{n!}~\rho^n\,,
\end{equation}
in field monomials $\rho^n$. Inserting \eq{poly} into \eq{flow} and solving for $\partial_t u=0$
leads to unique algebraic expressions for $\lambda_n$ with $n\neq 1$ as
functions of the mass term at vanishing field $u'(0)=\lambda_1 \equiv \mass$. Notice that we could have retained a vacuum  term $u(0)=\lambda_0$, whose explicit solution would then be of the form
\begin{equation}
\lambda_0=\frac{N}{3(1+\mass)}\,.
\end{equation}
However, this term has no influence whatsoever on the solution: changes in $\lambda_0$ leave the fixed point solution and universal scaling exponents unaffected, and therefore it can be set to zero from the outset.
(For other choices of $u(0)$, $\lambda_0$ is modified correspondingly.)
For the higher order polynomial couplings, we find 
\beq\label{recursiveN}
\lambda_{n+1}
= \frac{1}{2n+N} \bigg[ (1+m^2) \sum_{l=0}^{n-1} \Lambda_{n,l} 
+ \sum_{k=1}^{n-1} \sum_{l=0}^k \binom{n}{k} \Lambda_{k,l} \lambda_{n-k+1} \bigg] \, .
\eeq
where
\bea
\Lambda_{i,j} &=& \binom{i}{j}
[\delta_{j} + (2j+1)\lambda_{j+1}] 
(i-j-3) \lambda_{i-j}\,.
\eea
Resolving \eq{recursiveN} in terms of the sole free parameter $\mass$, we find
\begin{equation}\label{SmallFieldSolution}
\lambda_n=m^2(1+\mass)^n \ (N+2)^{-2} \ P_n(x,N)\,.
\end{equation}
Here, the functions $P_n(x,N)$ for $n\ge 2$ are recursively defined polynomials in
\begin{equation}\label{x}
x\equiv\frac{m^2}{N+2}
\end{equation}
of degree $n-2$ whose rational coefficients in $N$ may have poles only for negative even integer $N<-2$.\footnote{The fixed point solution \eq{SmallFieldSolution} are linked to the explicit solution
given for $N=1$ in (3.5) of \cite{Litim:2002cf} by
$\lambda_n\to(6\pi^2)^{n-1}\lambda_n$.}  Thus, the solution has a similar structural dependence on  $m^2$ as  the solution \eq{lambdam} established previously for the infinite-$N$ limit. The first few polynomials $P_n(x)$ read
 \bea
P_2&=&-2(N+2)\,,\quad
\nonumber\\
P_3&=&\
2\0{(N+2)}{N+4} +2x\, \0{(N+2)(5N+34)}{N+4}\,,\quad
\nonumber\\
P_4&=&
-24 x\,\0{(N+2)(N+14)}{(N+4)(N+6)} 
-24x^2\,\0{(N+2)(3N^2+44N+268)}{(N+4)(N+6)}\,,\nonumber\\
P_5&=&\ \
48
x^3
\,\0{(N+2)(99696 + 35848 N + 4248 N^2  + 370 N^3 + 13 N^4)}{(N+4)^2(N+6)(N+8)}
 \nonumber\\
&&+48 x^2
\,\0{(N+2)(6928 + 2016 N + 126 N^2  + 5 N^3)}{(N+4)^2(N+6)(N+8)}
\nonumber\\ && \label{solutionExpansion}
+96 x\,\0{(N+2)(3N+22)}{(N+4)^2(N+6)(N+8)}\,,
\eea
and similarly to higher order. A few comments are in order.
The coefficients in \eq{SmallFieldSolution}, \eq{solutionExpansion} are finite and well-defined for all $N$ except for even negative integers. A special role is taken by $N=-2$ 
 where finiteness of the fixed point solution requires finiteness for \eq{x} in the limit $N\to -2$, and for all even negative integer $N$ where some of the coefficients \eq{solutionExpansion}  become singular.
In the infinite-$N$ limit the polynomial couplings \eq{recursive}, \eq{lambdaB} are reproduced from \eq{SmallFieldSolution} using the appropriate rescaling with powers of $N$,
\begin{equation}\label{lambdaN}
\lambda_n\to \lim_{N\to \infty} \lambda_n(N)\,N^{n-3}\,,
\end{equation}
showing that the recursive relations and their solutions are smoothly connected via $N$ for all $N$.

\begin{figure*}[t]
\begin{center}
\includegraphics[width=.9\hsize]{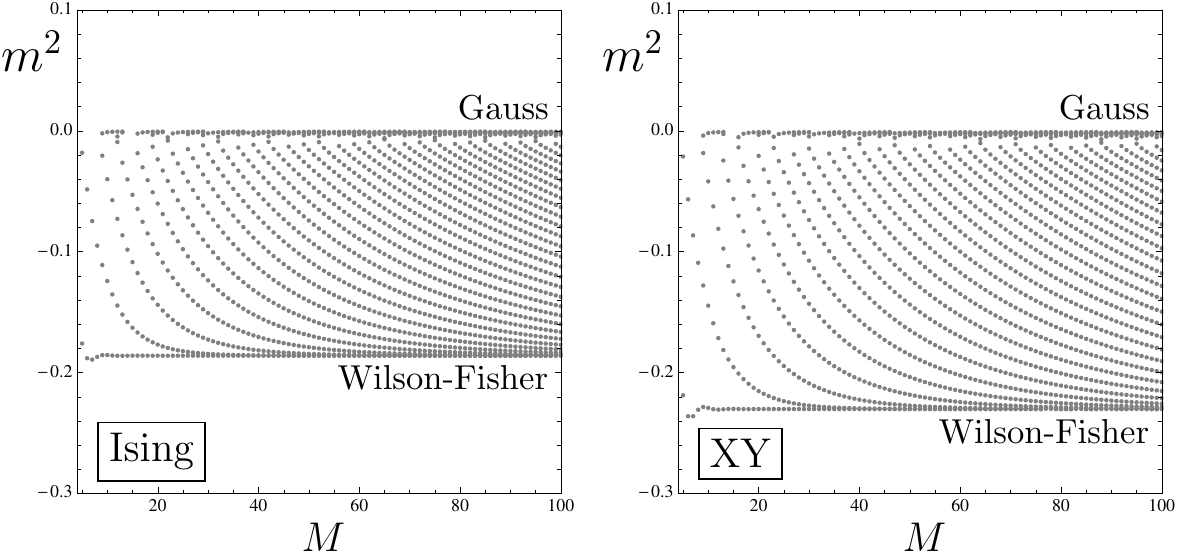}
\vskip.3cm
\caption{Shown are the non-trivial real roots $m^2$ of the auxiliary condition \eq{boundaryN}  as a function of the approximation order $M$ for the Ising ($N=1$, left panel) and the XY ($N=2$, right panel) universality classes, with fixed point couplings  given by the expressions \eq{SmallFieldSolution}.   With increasing order $M$, the values for admissible real roots accumulate  close to the  Gaussian \eq{mGauss} and  the Wilson-Fisher fixed points \eq{mWF1} and \eq{mWF2}, respectively. The convexity fixed point, not displayed in this graph, appears both as an exact solution and as an accumulation point.}
\label{pRoots}
\end{center}
\end{figure*}

Turning to the fixed point solutions, the algebraic solution \eq{SmallFieldSolution} displays the exact Gaussian fixed point 
 \begin{equation}\label{mGauss}
 \mass=0
 \end{equation} 
 which trivially entails the vanishing of all higher order couplings $\lambda_n\equiv 0$. 
 Similarly to the result in the infinite-$N$ limit \eq{convexInf}, the expressions \eq{SmallFieldSolution}  also display the 
 exact fixed point 
 \begin{equation}\label{-1}
 \mass=-1
 \end{equation}
 with $\lambda_n\equiv 0$ to all orders in the expansion. 
 The Wilson-Fisher fixed point corresponds to a specific value for $-1<m^2<0$, which is not directly fixed by the recursive solution and remains to be determined for all finite $N$ by some other mean. 
To that end, we may analyse the set of real roots of the auxiliary condition
\begin{equation}\label{boundaryN}
\lambda_{M+1}(m^2)=0\,.
\end{equation}
This auxiliary condition corresponds to the approximation where interactions terms up to $\lambda_{M}\rho^{M}$ are retained in \eq{poly}. Besides the (trivial) exact multiple roots at \eq{mGauss} and \eq{-1}, the auxiliary condition has in principle $M-2$ complex roots, and thus a total of 4752 roots for all $M$ between 4 and 99 inclusive. Among these,  we find a total of 1790 (1726) real roots  in the range $0>m^2>-1$ for the $N=1$ $(N=2)$ universality class, respectively. The non-trivial real roots are displayed in Fig.~\ref{pRoots}. Notice that the exact root $m^2=0$ of \eq{SmallFieldSolution} is not part of the plotted data. Still, with increasing $M$, the data displays two accumulation points corresponding to the Gaussian fixed point \eq{mGauss}, and to the Wilson-Fisher fixed point for the Ising universality class where
\begin{equation}\label{mWF1}
 \mass=-0.186\,064\cdots\,,
 \end{equation} 
and for the XY universality class $(N=2)$, where
 \begin{equation}\label{mWF2}
 \mass=-0.230\,185\cdots\,.
 \end{equation}
 We conclude that the accumulation points for the roots of the auxiliary condition \eq{boundaryN} provide  evidence  for the ``quantisation'' of fixed point solutions. Ths strategy then also allows for a good  determination of the physically relevant values for the parameter $m^2$.

Finally, we note that the field-dependent masses of the Goldstone modes $m^2(\rho)\equiv u'(\rho)$ and the radial mode $M^2\equiv u'(\rho)+2\rho u''(\rho)$ read
\begin{eqnarray}
m^2(\rho)&=& \mass -\frac{1}{2} \, \frac{\mass}{N+2}\, (1+\mass)^2\, \rho + \cdots \\
M^2(\rho) &=& \mass -\frac{3}{2} \, \frac{\mass}{N+2}\,(1+\mass)^2\, \rho + \cdots\quad
\end{eqnarray}
in the vicinity of small fields, using \eq{SmallFieldSolution}. At vanishing field both of these coincide, $M^2=m^2$. Away from it we observe that their ordering changes from $M^2>m^2$ for $\rho>0$ to $-1\le M^2\le m^2$ for $\rho<0$. Note also that $M^2$ stays strictly above $-1$ for all fields, in contrast to the mass of the `would-be' radial mode at infinite $N$.

\subsection{Large fields}

At large fields $\rho/\rho_0\gg1$, the potential approaches an infinite Gaussian fixed point
\beq
u_*'(\rho)=\gamma(N)\,\rho^2\,.
\eeq
All fluctuations are suppressed for $1/\rho\to 0$, in full analogy to the results found in Sec.~\ref{sec:Ninfty_realfields}. 
We therefore may follow the strategy given there and expand the Wilson-Fisher fixed point about the infinite Gaussian fixed point
\begin{eqnarray}\label{LargeFieldsI}
u'&=& \gamma(N)\,\rho^2\left[1+\sum_{n=1}^{\infty} ~ \gamma_n(N) ~ \rho^{-n}\right]\,.
\end{eqnarray} 
The iterative solution determines the coefficients $\gamma_n(N)$ of the remaining field monomials as unique algebraic functions of the leading-order coefficient $\gamma(N)$ and $N$. 
Explicitly, we find
\begin{eqnarray}
\label{largefieldC}
\gamma_5(N) &=& -\frac{2}{5}(N-\s045) \frac1{\gamma(N)^2}\,, \nonumber\\
\gamma_7(N) &=& \ \  \, \frac{4}{7}(N -\s0{24}{25})\frac1{\gamma(N)^3}  \nonumber\\
\gamma_9(N) &=&-\frac{2}{3}(N-\s0{124}{125}) \frac1{\gamma(N)^4}\\ \nonumber
\gamma_{10}(N) &=& -\frac{7}{25} (N-\s065) (N -\s045)\frac1{\gamma(N)^4}
\end{eqnarray}
for the first few coefficients. Identifying the global WF solution at finite $N$ corresponds to
determining the remaining free parameter $\gamma(N)$ per universality class. 

Comparing with the infinite-$N$ limit \eq{lambdaC} we note that the radial mode only introduces mild modifications in the structure of the solution. Substituting
\begin{equation}
\gamma(N)=\frac{\gamma}{N^2}\,,\quad \gamma_n(N)=\gamma_n N^{n}
\end{equation}
into the coefficients \eq{largefieldC}  and taking the infinite-$N$ limit, we confirm that these fall back on \eq{lambdaC}, establishing that the recursive solutions at large fields are smoothly connected for all $N$.

\begin{figure}[t]
\begin{center}
\includegraphics[width=.6\hsize]{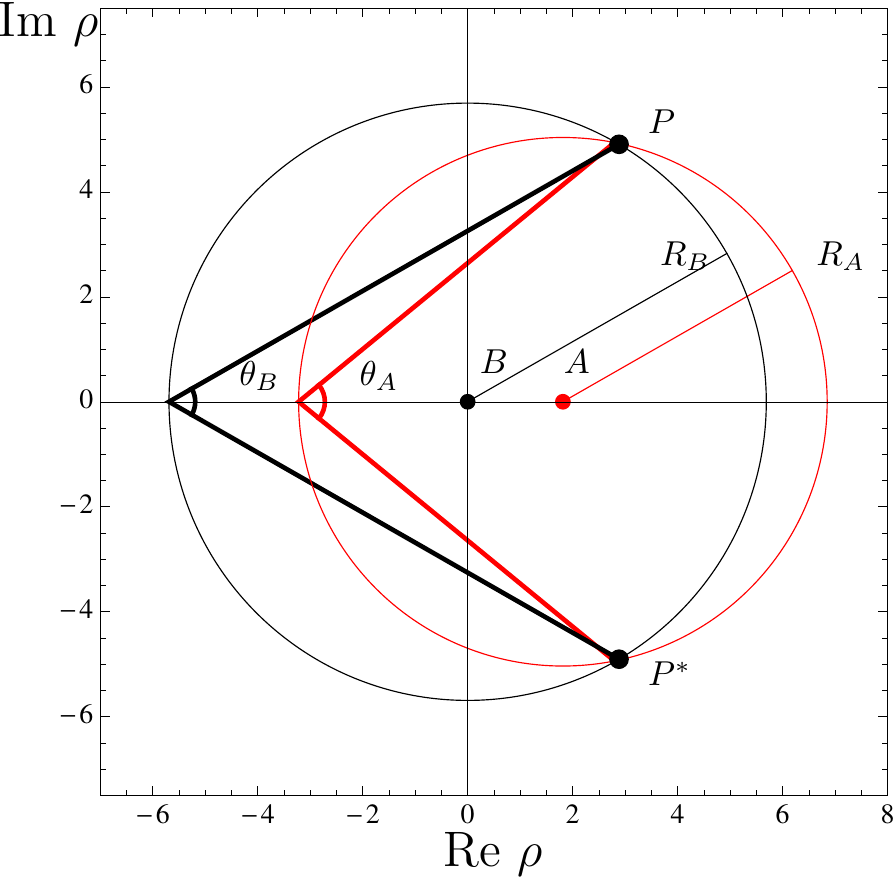}
\vskip.3cm
\caption{Singularity of the Wilson-Fisher fixed point  (Ising universality class) in the complex field plane. Shown are the location of the convergence-limiting poles $P$ and $P^*$ (black dots)   in the complexified field plane, the angles $\theta$  and the radii of convergence $R$ for the expansions about the minimum $(A$, red) and and vanishing field $(B$, black).}
\label{pIsing}
\end{center}
\end{figure}

\subsection{Singularities in the complex field plane}
We may now exploit our findings to estimate the location of convergence-limiting poles in the complex field plane. To that end, we consider the expansions $A$ and $B$ at the example of the Ising universality class $N=1$. We fix the remaining free parameters  $(\rho_0,\lambda)$ for the expansion $A$ about  the local minimum, and the parameter $m^2$ for the expansion $B$ about vanishing field  
using input from numerical studies \cite{Bervillier:2007rc},
\begin{equation}\label{input}
\begin{array}{rl}
m^2&=-0.186\,064\,249\,470\,314\,52\\[.5ex]
\rho_0&=\ \ \,1.814\,898\,403\,687 \\[.5ex]
\lambda&=\ \ \,0.126\,164\,421\,218\,.
\end{array}
\end{equation}
The numbers \eq{input}, together with the first 50 coefficients \eq{lambdasNA} for the expansion $A$ provides us with estimates for the radius of convergence and the nature of the singularity by using the Mercer-Roberts test for the series \eq{minifinin}. We find 
\begin{equation}\label{IsingA}
\begin{array}{rl}
R_A&=5.03(5)\\[.5ex]
\theta_A&=78.74(1)^\circ\\[.5ex]
\nu_A&=1.5(05)\,.
\end{array}
\end{equation}
The locaton of the pole in the complex field plane is displayed in Fig.~\ref{pIsing}. 

For the expansion $B$, we use \eq{input} to obtain the first 50 expansion coefficients \eq{SmallFieldSolution} numerically, and deduce the radius of convergence of the series \eq{poly}  from the Mercer-Roberts test as 
\begin{equation}\label{IsingB}
\begin{array}{rl}
R_B&=5.69(0)
\\[.5ex]
\theta_B&=59.5(45)^\circ\\[.5ex]
\nu_B&=1.49(92)\,.
\end{array}
\end{equation}
The result is displayed in Fig.~\ref{pIsing}, where the locations of the poles $P$ and $P^*$ are indicated by dots. Both expansions appear to be controlled by the same singularity, as can be seen from  Fig.~\ref{pIsing} and the agreement of the nature of the singularity $\nu_A\approx \nu_B$ from \eq{IsingA} and \eq{IsingB}. It is interesting to note that the nature of the singularity has changed from a square-root type behaviour \eq{nu12}  at infinite $N$ to a less pronounced $\nu =\s032$ behaviour. Consequently, at finite $N$, the sextic interaction $u'''$ is the first one to display a singularity in the complex field plane
\begin{equation}\label{u3pole}
u'''(\rho)\sim\frac{1}{\sqrt{\rho-\rho_s}}
\end{equation}
in the vicinity of some $\rho_s$, in contrast to the result \eq{u2pole} at infinite $N$. We conclude that the expansions $A$ and $B$ for the Ising universality class continue to be controlled by one and the same  pole in the complex field plane, similar the infinite-$N$ limit, except that the precise location of the pole and the nature of the singularity depend on the universality class.

\subsection{Imaginary fields}\label{imagine}

The fluctuations of the radial mode 
contribute the term $(3u''+2\rho u''')I'[u'+2\rho u'']$ to the RG flow \eq{eq:flowprime} for the effective action.
Using an optimised cutoff \eq{I}, the radial contribution reads 
\begin{equation}\label{radial}
-\frac{3u''+2\rho\,u'''}{(1+u'+2\rho\,u'')^2}\,.
\end{equation}
This term modifies the analyticity structure of fixed 
point solutions, in particular for negative $\rho$.
This can be understood as follows. Let us suppose that
\begin{equation}\label{expansion}
1+u'=\frac{a}{\sqrt{-\rho}}+\frac{b}{\rho}+{\rm subleading}\,,
\end{equation}
in the limit of large negative $\rho$. This becomes the leading-order solution in the large-$N$ case provided $a=0$. The expansion ansatz \eq{expansion} satisfies the fixed point equation \eq{eq:flowprime} in two ways, with either $a=0$ or $a\neq0$. In the first case, inserting \eq{expansion} into \eq{eq:flowprime}, we find
\begin{equation}\label{coeffLN}
a=0\,, \quad b=\frac{2-N}{2}\,.
\end{equation}
Then $\partial_t u'=0$ is fulfilled to leading order because the term $(-2u')$ 
in \eq{eq:flowprime} is cancelled jointly by the Goldstone and by the 
radial mode. The subsequent iteration  then determines all subleading 
corrections exactly as we did in the infinite-$N$ case.  
Notice that positivity of $(1+u')$ requires that $b/\rho>0$, meaning that $b<0$. 
On the other hand, for $a\neq0$ there is a new branch of expansions available, 
starting off with
\begin{equation}\label{coeffFN}
a>0
\,,\quad b=\frac{1}{2}>0\,.
\end{equation}
In this case, the  term $(-2u')$ in \eq{eq:flowprime} is cancelled only by the radial mode, and the Goldstone contribution has become subleading.

Next, we consider the function $(1+u'+2\rho u'')$, the argument in the denominator of the radial contribution \eq{radial}. Finiteness of the full flow requires that the pole at $1+u'+2\rho u''=0$ cannot be crossed, and hence both 
\beq\label{pos}
\begin{array}{rcl}
0&<&1+u'\\[.5ex]
0&<&1+u'+2\rho u''
\end{array}
\eeq 
must hold for all finite $\rho$. (Notice that at infinite $N$, the second condition is neither required nor satisfied, see Fig.~\ref{Amplitude}.) Evaluating \eq{pos} in the limit $\rho\to-\infty$, using \eq{expansion}, we find
\begin{equation}\label{expansion2}
1+u'+2\rho u''= \frac{b}{(-\rho)}+{\rm subleading}\,.
\end{equation}
We note that the asymptotic behaviour is independent of $a$. This comes about because the operator $(1+ 2\rho\partial_\rho)$ has a ``zero mode'' $\propto|\rho|^{-1/2}$. We conclude that the parameter $b$ must obey $b>0$, which applies for the expansion \eq{expansion2}. Consequently, an asymptotic expansion \eq{expansion} with \eq{coeffLN} cannot be achieved for $N>2$. For our purposes, this excludes the expansion starting as \eq{coeffLN} and imposes an expansion starting with \eq{coeffFN}.  We are therefore lead to the series
 \begin{equation}\label{AnsatzImaginaryN}
u'(\rho)=-1+\sum_{m=1}^{\infty}\sum_{n=0}^{m-1}
\zeta_{m,n}\,\left(\sqrt{-\rho}\right)^{-m} \,\ln^n(\sqrt{-\rho})\,,
\end{equation}
which should be compared with its counterpart \eq{AnsatzImaginaryLN} at infinite $N$. Similarly to \eq{AnsatzImaginaryLN}, we observe the appearance of subleading logarithmic terms. The main difference is that the analyticity structure has changed owing to the radial mode.
 
We have determined the first few hundred expansion coefficients \eq{AnsatzImaginaryN}  after solving the fixed point condition recursively for all couplings. We find that the series  can be expressed in terms of two free expansion parameters $\zeta$ and $\bar\zeta$,
\begin{equation}\label{xis}
\begin{array}{rl}
\zeta_{1,0}&\equiv\zeta\\[.5ex] 
\zeta_{4,0}&\equiv\bar\zeta\,.
\end{array}
\end{equation}
In terms of these, the first few non-vanishing coefficients are
\begin{eqnarray}
\zeta_{2,0}&=&-\012\,,\nonumber\\
\zeta_{3,0}&=&-\058\zeta -\0{N-1}{8\zeta}\,, \nonumber\\
\zeta_{4,1}&=&\ \  \, \014\,,  \\
\zeta_{5,1}&=&\ \  \, \0{15}{16}\,\zeta+\0{3(N-1)}{16\zeta}\,,   \nonumber
\end{eqnarray}
and similarly to higher order. Beyond this order, the coefficients $\zeta_{m,n}$ become functions of both $\zeta$ and $\bar\zeta$, eg.~$\zeta_{5,0}=-\s0{5}{128}\,\zeta\,(9-96\,\bar\zeta-50\,\zeta^2)$ for $N=1$. To identify the Wilson-Fisher fixed point amongst these candidate fixed points, the free parameters \eq{xis} need to be determined by other means.

\begin{table*}
\begin{center}
\begin{tabular}{c|ccccc}
\rowcolor{LightGreen}
\toprule
\rowcolor{LightGreen}
&&&&&\\[-2.5mm]
\rowcolor{LightGreen}
\bf expansion&${}\quad \quad  \bm A\quad \quad$&${}\quad \quad \bm B\quad \quad$&${}\quad \quad \bm C\quad \quad$&${}\quad\quad  \bm D\quad \quad$&
\\ 
\rowcolor{LightGreen}
${}\quad$expansion point$\quad$&${}\quad \rho=\rho_0\quad$&${}\quad \rho=0\quad$&$1/\rho=0^+$&$1/\rho=0^-$&\multirow{-2}{*}{{}$\quad \quad \bm N\quad \quad$}
\\ 
\midrule
\rowcolor{LightGray}Goldstone modes only&none&$m^2$&$\gamma$&$\zeta$& infinite\\
Goldstone and radial modes&$(\rho_0,\ \lambda)$&$m^2$&$\gamma$
&$(\zeta,\ \bar\zeta)$&finite\\
\rowcolor{LightGray}
radial mode only&$(\rho_0,\ \lambda)$&$m^2$&$\gamma$
&$(\tau,\ \bar\tau)$& $1$\\
\bottomrule
\end{tabular}
\caption{Free parameters arising as part of the exact recursive fixed point solutions derived in the main text.}
 \label{tParameters}
\end{center}
\end{table*}

\subsection{Ising universality}

Although the case of the Ising universality is covered by our previous discussion, it is useful to discuss the case $N=1$ from a different angle. In the Ising case, the Goldstone modes are absent throughout, and we are left with only the radial mode. 
We make two observations.
Firstly, we notice that the exact recursive solution \eq{recursiveN}  for the polynomial couplings \eq{poly}  simplifies in the Ising limit. In terms of $\lambda_1\equiv m^2$, we find
\beq\label{recursive1}
\lambda_{n+1}=
\left( \frac{1+m^2}{1+2n} \right)\bigg[(n-3)\lambda_n
+ \sum_{k=0}^{n-1} \binom{n}{k} (2k+1)(n-k-3)\lambda_{2k+1}\lambda_{n-k}\bigg]\,.
\eeq
This result should  be compared with the result in the infinite $N$ limit \eq{recursive}. As is evident from the explicit form \eq{recursive1}, both the Gaussian and the convexity fixed point are exact solutions to each and every order in the polynomial expansion of the effective potential around vanishing field.
Secondly, we notice that  the RG flow can solely be formulated in terms of the field-dependent radial mass
\begin{equation}\label{w}
w(\rho)=u'(\rho)+2\rho u''(\rho)\,,
\end{equation}
leading to
\begin{equation}\label{w1}
\partial_t w=-2w+\rho w'-\frac{w'+2\rho w''}{(1+w)^2}+\frac{4\rho(w')^2}{(1+w)^3}\,.
\end{equation}
Notice that this change of variables $u'\to w$ also implies that we are no longer sensitive to one parameter related to the `zero-mode' of $1+2\rho\partial_\rho$ when reconstructing $u'$ from $w$. 

From the structure of the flow \eq{w1}, we conclude that an algebraic solution around vanishing field and the minimum of $w$ will supply us with algebraic expressions for all couplings in dependence of one and two free paramters, respectively. Equally, an asymptotic expansion about large real fields will provide us with a one-parameter family of solutions. 

In the regime of purely imaginary fields where $0<1+w\ll 1$, the flow allows for an asymptotic expansion of the form
\begin{equation}\label{w1ansatz}
w(\rho)=-1+\sum_{n=2}^{\infty}\sum_{m=0}^{n-1}
\tau_{n,m}(\sqrt{-\rho})^{-n} \ln^m(\sqrt{-\rho})\,.
\end{equation}
of which we have computed the first hundred coefficients. 
Expanding \eq{w1} in the `operator basis'  \eq{w1ansatz} we find the couplings $\tau_{m,n}$ recursively. Here, as opposed to the expansion in \eq{AnsatzImaginaryN}, there is no term $\propto (-\rho)^{-1/2}$ as it corresponds to a zero mode of the differential operator $\partial_\rho+2\rho\partial^2_\rho$, see \eq{w}.
Adopting the same strategy as before, all couplings can be expressed as functions of two free parameters
\begin{equation}
\begin{array}{rl}
 \tau&=\tau_{3,0} \\[.5ex]
 \bar\tau& =\tau_{4,0}\,.
 \end{array}
 \end{equation}
 In terms of these, the first few non-vanishing coefficients are 
\begin{eqnarray}
\tau_{2,0}&=&\ \  \, \0{1}{2}\,,\quad
\nonumber \\
\tau_{4,1}&=&-\0{3}{4}\,,
\quad 
\nonumber \\
\tau_{5,0}&=& \ \  \, \0{1}{8} (7 \tau -32 \tau^3 +32 \tau \bar\tau )\,,\quad
\nonumber\\
\tau_{5,1}&=&- 3\tau\,, \label{zetaN1} \\
\tau_{6,0}&=&\ \  \,  \0{1}{96} 
(-17 + 336 \tau^2 -768 \tau^4 +32 \bar\tau 
+384 \tau^2 \bar\tau +192 \bar\tau^2 )\,,\nonumber 
\\
\tau_{6,1}&=& - \0{1}{4} (1+12 \tau^2 -12 \bar\tau)\,, \nonumber \\
\tau_{6,2}&=&\ \  \,  \0{9}{8}\,,  \nonumber
\end{eqnarray}
and similarly to higher order. As a consistency check, we insert the result given in  \eq{AnsatzImaginaryN}  into \eq{w}. The expansion \eq{zetaN1} is thereby recovered  by subtituting
\begin{equation}
\zeta = \frac{4}{5}\,\tau ~~~\textrm{and}~~~
\bar{\zeta} = \frac{1}{12}-\frac{\bar\tau}{3} 
\end{equation}
and by setting $N=1$. We conclude that either way is practibable to establish the asymptotic expansion. The formulation used in this subsection centrally differs due to the absence of a leading $1/\sqrt{-\rho}$ contribution, which in turn is due to the absence of Goldstone modes.

\subsection{Discussion}

We summarise the main results at finite $N$ and compare with those at infinite $N$. In either case, fixed point solutions are found via the systematic expansions about small, large and imaginary fields. Interestingly, the expansion about vanishing and asymptotically large fields remain qualitatively the same, being controlled by a single free parameter irrespective of the universality class $N$.  
On the other hand, the expansions about the potential minimum and about large and purely imaginary fields have become more complex due to the fluctuations of the radial mode as soon as $N$ is finite. Here, the recursive solutions depend on two rather than none or one free parameter, respectively; see Tab.~\ref{tParameters} for an overview. 
Another important difference between finite and infinite $N$ arises in the expansion  $D$ about large imaginary field. Here, and unlike for the expansions $A$, $B$ and $C$,  the fluctuations of the radial mode  imply that the infinite-$N$ limit is not continuously connected to the results for any finite $N$.
Further criteria such as the vanishing of higher order couplings \eq{boundaryN} can  be invoked to uniquely determine the remaining free parameters and the corresponding fixed point solutions at finite order in the approximation.  At infinite $N$, the Wilson-Fisher fixed point then appears  as a unique isolated solution, as illustrated in Fig.~\ref{pRootsInf}. At finite $N$,  the physical solution appears as  an accumulation point, owing to the presence of the radial field fluctuations, see Fig.~\ref{pRoots}. In either case, these patterns are sufficient to identify fixed points reliably within polynomial expansions. 

We also have established that the convergence of the various expansions is controlled by singularities of the Wilson-Fisher fixed point in the complex field plane, related to poles in the quartic (infinite $N$) or sextic (finite $N$) scalar self-interaction. On the level of the RG flow for the field dependent mass, the singularity relates to a non-analytical dependence on the complexified field of the form
\beq \label{singular}u'(\rho)\propto \left(\sqrt{\rho-\rho_s}\right)^{-n}\eeq 
with $n=1\, (n=3)$ for infinite (finite) $N$, respectively, showing that certain $n$-point functions at the fixed point  at vanishing external momenta display singularities for specific  points  in the complex field plane (outside the physical domain). In either case, the local expansion coefficients are sufficient to determine the location of convergence-limiting poles, see Fig.~\ref{pIsing}. Based on the results at infinite $N$, it is  conceivable that  key characteristics of the Wilson-Fisher fixed point including its scaling exponents can be reliably deduced from local approximations, even in the case of finite $N$. This viewpoint is  supported by the size of the radius of convergence, which we have determined in \eq{IsingA} and \eq{IsingB} for $N=1$. It would be useful to confirm these results for all $N$  \cite{numerical}.  

\subsection{Higher orders, resummations, and conformal mappings}

We close with a few remarks regarding natural extensions of our work. Firstly, 
 it is straightforward to apply our technique beyond the local potential approximation,
leading to additional field-dependent functions besides the effective potential. For each of these, the polynomial expansions can be performed, and  the convergence-limiting singularities can be localised in the complex plane using the methods developed here. If anomalous dimensions remain small, our leading order results should receive only mild corrections quantitatively.
 We expect that insights into the singularity structure at higher order may also help clarify the notorious convergence of the derivative expansion.

Secondly, our results may be enhanced in combination with  suitable resummations such as Pad\'e, thereby extending the domain of validity of polynomial fixed point solutions \cite{Bervillier:2007tc,Jakovac:2013jua,Falls:2016wsa}. In fact, the exact recursive expressions for the fixed point couplings provides crucial input for any resummation scheme. It has already been observed that suitably adapted resummations extend the radius of convergence for curvature expansions in quantum gravity \cite{Falls:2016wsa}. These observations can straightforwardly be adapted for the theories studied here.

Finally, our results may  be combined with the technique of conformal mappings which  has been developed for functional flows  in  \cite{Bervillier:2008an}. Central input for this technique is the precise knowledge of singularities in the complex field plane closest to the origin, most notably their distance $R$ from the origin 
and the  opening angle $\alpha\cdot \pi$ of the corresponding angular section in the complex field plane such as those shown in  Figs.~\ref{pZero} and \ref{pIsing}. 
Then, a change of variables from $u(\rho)$ onto $\tilde u(w)=u(\rho(w))$  according to the conformal map 
\beq \label{conformal}
\begin{array}{rl}
w&
\displaystyle
=\frac{(1+\rho/R)^{1/\alpha}-1}{(1+\rho/R)^{1/\alpha}+1}\,,
\end{array}
\eeq
together with mild assumptions for the large-field asymptotics, offers a flow for $\tilde u(w)$ whose expansions in small $w$ should converge in the entire disc   $|w|<1$  \cite{Bervillier:2008an}. When re-expressed in terms of the original variables, this leads to an enhanced domain of validity, covering  all $\rho$ within the entire angular section defined by the opening angle, and provided that no further singularities arise within the section. The parameter $(R,\alpha)$, central input for conformal mappings \eq{conformal}, relate as
\beq
R=R_A\,,\quad \alpha=\theta_A/\pi
\eeq
to the radius of convergence $R_A$ and the  opening angle $\theta_A$ of the polynomial expansion $A$ as developed here, see \eq{radiusA}, \eq{Aexact},  and \eq{IsingA}. It would thus seem promising to combine conformal mappings with our findings including at higher orders in the derivative expansion.

\section{\bf Conclusions}\label{Conclusion}

  We have put forward ways to solve $O(N)$ symmetric scalar field theories analytically  in the vicinity of  interacting  fixed points of their renormalisation group flow.  A main novelty are explicit recursive relations for couplings at small, large, or imaginary field, offering access to all scaling solutions of the theory. 
At finite polynomial order, physical fixed points appear either as unique isolated solutions or as accumulation points. In either case local fixed points are extended to global ones by matching additional parameter. 
The  accurate knowledge of the fixed point in the large-field region is relevant for the global scaling solution, but much less so for the determination of scaling exponents. For most practical purposes, we conclude that local polynomial approximations of the renormalisation group is sufficient to deduce scaling exponents \cite{Litim:2002cf}. 

We also found that derivatives of effective actions at fixed points genuinely display singularities in the complexified field plane \eq{singular}, away from the physical region. 
In the limit of infinite $N$,  exact analytical results for all singularities have been  provided. At finite $N$, we have put forward a methodology to deduce radii and location of singularities 
from local expansions. 
Singularities in the complex field plane impact on the physical solution in a number of ways. Most notably, they control the radius of convergence for polynomial approximations of the effective action. Global scaling solutions then follow from local ones as soon as the radii of convergence of the small- and large-field expansions overlap, which is the case for sufficiently large $N$. It will be interesting to complement this analytical study with  resummations \cite{Falls:2016wsa} and confomal mappings \cite{Bervillier:2008an} to further extend the radius of convergence, or with numerical tools to accurately determine the global fixed points and salient theory parameters, most notably for finite and small $N$ \cite{numerical}.

Some of the structural insights  should also prove useful for theories where global fixed point solutions are presently out of reach, e.g.~quantum gravity \cite{Benedetti:2012dx,Dietz:2012ic,Falls:2013bv,Falls:2014tra}. There, the recursive nature of fixed point couplings  has already been observed up to high polynomial order \cite{Falls:2013bv,Falls:2014tra}. It would then seem promising to investigate these theories from the viewpoint advocated here. 
\\[2ex]

\centerline{\bf Acknowledgements}
${}$\\
We thank  Andreas J{\"u}ttner  for discussions, comments on the manuscript, and collaboration on a related project. Some of our results have been presented at the 3rd UK-QFT workshop (Jan 2014, U Southampton). This work is supported by the Science and Technology Facilities
Council (STFC) under grant number ST/L000504/1.

\appendix
\renewcommand{\thesection}{{\bf \Alph{section}}}

\section{\bf Radius of convergence}\label{AppA}

We often have to estimate the radii of convergence of series of the form
\beq\label{A1}
f(x)=\sum_m\lambda_m x^m\,.
\eeq
Provided the signs of the expansion coefficients follow a simple pattern such as having all the same, or alternating signs $(+-)$, standard tests of convergence including the ratio test or the root test are applicable. Here, we encounter series which often show a more complex sign pattern such as $(++--)$ where the standard tests fail, or at best provide a rough estimate for the radius of convergence. 
Therefore, we adopt a method by Mercer and Roberts \cite{Mercer:1990} devised for series which display a more complex pattern. The technique is motivated from the expansion of functions on the real axis which display a pair of complex conjugate poles in the complex plane such as 
\beq\label{example}
f(x)=\left(1-\frac{x}{Re^{i\theta}}\right)^\nu+\left(1-\frac{x}{Re^{-i\theta}}\right)^\nu
\eeq
for  $R>0$, $0\leq \theta\leq \pi$, and for any $\nu$ which is neither zero nor a positive integer ($\nu\notin Z_0^+$). The function \eq{example} has a pair of singularities  of nature $\nu$ at $x=R e^{\pm i \theta}$. Its Taylor expansion is given by \eq{A1} with
\beq\label{la}
\lambda_n=
2(-1)^n\left(\begin{array}{c}\nu\\n\end{array}\right)\frac{\cos(n\,\theta)}{R^n}
\eeq
for $|x|<R$. For large $n$, the sign of the coefficients is determined by the phase. For example, a singularity close to the imaginary axis with phase close to $\theta\approx \pi/2$ would imply the sign pattern $(++--)$. Returning now to a generic series \eq{A1}, and under the assumption that the large-$n$ behaviour can be modeled by \eq{la}, we can write every 4-tuple of coefficients in the form
\beq\label{series}
\lambda_m=\frac{A_n}{(R_n)^m}\cos(m\theta_n+\xi_n)\,.
\eeq
The four parameters $A_n$, $R_n$, $\theta_n$ and $\xi_n$ appearing in \eq{series}  can then be determined from any 4-tuple $(\lambda_{n-2},\lambda_{n-1},\lambda_{n},\lambda_{n+1})$. For $R_n$ and the angle $\cos\theta_n$, one finds
\bea\label{Rn}
{R^2_n}&=&\frac{\lambda_{n-1}^2-\lambda_{n-2}\lambda_{n}}{\lambda_n^2-\lambda_{n-1}\lambda_{n+1}}\\
\label{cos}
\cos\theta_n&=&\frac12\left(
\frac{\lambda_{n-1}}{\lambda_n\,R_n}
+\frac{\lambda_{n+1}\,R_n}{\lambda_{n}}\right)\,.
\eea
For large $n$, the asymptotic behaviour is deduced by inserting the model coefficients \eq{la} into \eq{Rn} and \eq{cos}, leading to
\bea
&&\frac{R}{R_n}=\left(1-\frac{\nu+1}{n}\right)
\cdot\left(1+\frac{\nu+1}{2n^2}\frac{\sin(2n-1)\theta}{\sin\theta}\right)
\cdot\left(1+{\cal O}\left(\frac{1}{n^3}\right)\right) \,.
\eea
For the angles, one finds
\bea
&&\frac{\cos\theta_n}{\cos\theta}= 
1+
\frac{\nu+1}{n^2}\left(1-\frac{\cos(2n-1)\theta}{\cos\theta}\right)
+{\cal O}\left(\frac{1}{n^3}\right)\,.
\eea
These expressions in the limit $1/n\to 0$ have been used throughout this paper to determine the radius $R$, the angle $\theta$, and the nature $\nu$ of the convergence-limiting singularity of series of the form \eq{A1}. The technique fails provided that \eq{Rn} becomes negative, or the absolute value of \eq{cos} larger than one.

\bibliography{biblio2_modif}
\end{document}